\documentclass[twoside,11pt]{article}

\usepackage{blindtext}

\usepackage{amsmath}
\usepackage{enumitem}
\usepackage{subcaption}
\usepackage{float}
\usepackage{tabularx}
\usepackage{longtable}
\usepackage{tikz-cd}
\usepackage{xurl}
\usepackage{jmlr2e}

\newcommand{\Var}{\mathrm{var}}

\newcommand{\tr}{\mathrm{tr}}
\newcommand{\argmin}{\mathop{\mathrm{argmin}}}

\def\T{{ \mathrm{\scriptscriptstyle T} }}
\newcommand{\p}{\mathrm{pr}}
\newcommand{\f}{Fr\'{e}chet }

\usepackage{lastpage}
\jmlrheading{23}{2022}{1-\pageref{LastPage}}{6/22; Revised
	11/22}{11/22}{22-0681}{Yidong Zhou and Hans-Georg M\"{u}ller}

\ShortHeadings{Network Regression with Graph Laplacians}{Zhou and M\"{u}ller}
\firstpageno{1}

\begin{document}

\title{Network Regression with Graph Laplacians}

\author{\name Yidong Zhou \email ydzhou@ucdavis.edu \\
	\addr Department of Statistics\\
	University of California\\
	Davis, CA 95616, USA
	\AND
	\name Hans-Georg M\"{u}ller \email hgmueller@ucdavis.edu \\
	\addr Department of Statistics\\
	University of California\\
	Davis, CA 95616, USA}

\editor{Michael Mahoney}

\maketitle

\begin{abstract}
Network data are increasingly available in various research fields, motivating statistical analysis for populations of networks,  where a network as a whole is viewed as a data point. The study of how a network changes as a function of covariates is often of paramount interest. However, due to the non-Euclidean nature of networks, basic statistical tools available for scalar and vector data are no longer applicable. 
This motivates an extension of  the notion of regression to the case where responses are network data. Here we propose to adopt conditional \f means implemented as M-estimators that depend on weights derived from both global and local  least squares regression, extending the \f regression framework  to networks that are quantified by their graph Laplacians. The challenge is to characterize the space of graph Laplacians  to justify the application of \f regression. This characterization then leads to asymptotic rates of convergence for the corresponding M-estimators by applying empirical process methods. We demonstrate the usefulness and good practical performance of the proposed framework with simulations and with network data arising from resting-state fMRI in neuroimaging, as well as New York taxi records.
\end{abstract}

\begin{keywords}
  \f mean, graph Laplacian, neuroimaging, power metric, sample of networks
\end{keywords}

\section{Introduction}
\label{sec:intro}
Advances in modern science have led to the increasing availability of large collections of networks where a network is viewed as a fundamental unit of observation. Such data are encountered, for example, in the analysis of brain connectivity \citep{fornito2016fundamentals} where interconnections among brain regions are collected for each patient under study, and traffic mobility \citep{von2009public}, where volumes of traffic among transit stations are recorded for each day. Several recent studies focus on the analysis of collections of networks. For example, a geometric framework for inference concerning a population of networks was introduced and complemented by asymptotic theory for network averages in \citet{gine:17} 
and \citet{kolaczyk2020averages}. A similar framework for studying populations of networks, where the graph space is viewed as the quotient space of a Euclidean space with respect to a finite group action,  was studied in \citet{calissano2020populations} and a flexible Bayesian nonparametric approach for modeling the population distribution of network-valued data in \citet{durante2017nonparametric}. Recently various distance-based methods for collections of networks have also been proposed \citep{donnat2018tracking, josephs2020bayesian, wills2020metrics, lunagomez2021modeling}.

A challenging and commonly encountered problem is to model the relationship between network objects and one or more explanatory variables. Such regression problems arise, for instance, when one is interested in varying patterns of brain connectivity networks across covariates of interest, such as age and cognitive ability of subjects. As the elderly population increases, an emerging research topic is brain aging and especially age-related cognitive decline \citep{ferreira2013resting, sala2014changes, sala2015reorganization}. With  advances in neuroimaging techniques, and specifically fMRI, one is able to model the human brain as a network, where nodes correspond to anatomical regions and edges to the functional or structural connections among them. Normal brain aging can then be  studied by network-response regression, where the brain connectivity network of a subject is the response and the subject's age the predictor. This setting is different from the time-series case where the emphasis is on modeling a sequence of  
networks \citep{zhu2017network, kim2018review, jian:20, dubey2021modeling, wang2021high}.

Although various approaches have been proposed for regression with non-Euclidean responses, such as \citet{jain2016geometry}, \citet{cornea2017regression} or \citet{dai2018principal} among others, there are only very few studies  on network-response regression. Matrix representations of networks, such as adjacency matrices and graph Laplacians, are commonly used characterizations of the space of networks. \citet{wang2017bayesian} proposed the Bayesian network-response regression with a single scalar predictor by vectorizing adjacency matrices. This approach is restricted to binary networks and is computationally intensive due to the MCMC procedure involved in the posterior computation. Another approach for network-response regression is the tensor-response regression model \citep{zhang2018generalized, chen2021statistical}, where one models adjacency matrices locally by extending generalized linear models to the matrix case and imposes low-rank and sparse assumptions on the coefficients. A general framework for the statistical analysis of populations of networks was developed by embedding the space of graph Laplacians in a Euclidean feature-space, where linear regression was applied  using extrinsic methods \citep{severn2019manifold}. Nonparametric network-response regression based on the same framework was proposed subsequently by adopting Nadaraya-Watson kernel estimators \citep{severn2021non}. However, embedding methods suffer from losing much of the relational information due to the non-Euclidean structure of the space of networks and assigning nonzero probability to points in the embedding space that do not represent networks. Another practical issue in this context is  the need to estimate a covariance matrix which has a very large number of parameters. Based on the  adjacency matrix of a network, \citet{calissano2022graph} proposed a network-response regression model by implementing linear regression in the Euclidean space and then projecting back to the ``graph space'' through a quotient map \citep{calissano2020populations}. This model is widely applicable for various kinds of unlabeled networks, but for the regression case there is no supporting theory and this approach may not be suitable for labeled networks, which are prevalent in applications, for example  brain connectivity networks \citep{fornito2016fundamentals}.  

\begin{figure}[t]
	\centering
	\includegraphics[width=0.9\linewidth]{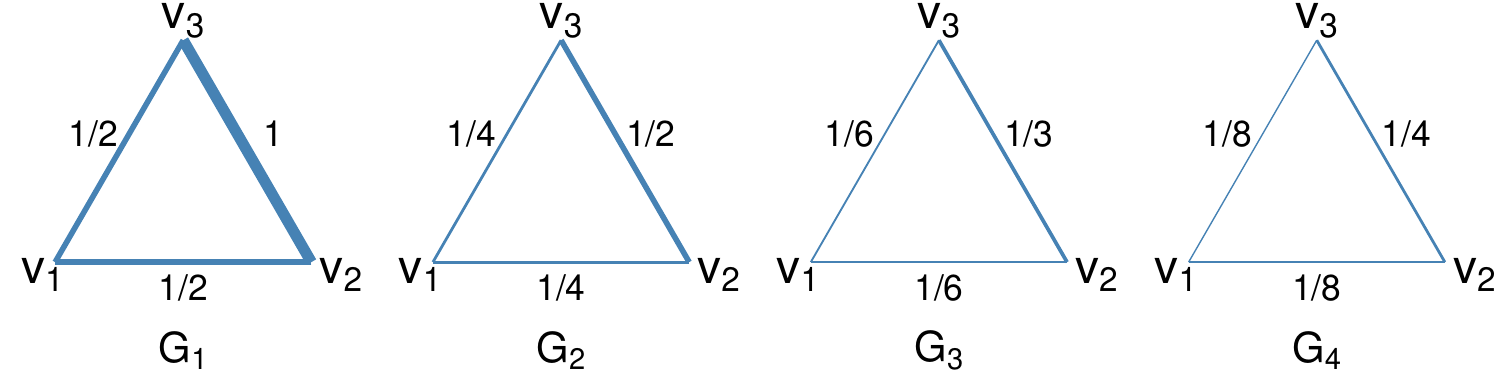}
	\caption{A toy example where networks $G_1, G_2, G_3, G_4$, each with 3 nodes, are independently observed, associated with one dimensional covariates $X_1=2, X_2=4, X_3=6, X_4=8$. The weight of each edge is marked beside the corresponding edge.}
	\label{fig:exm}
\end{figure}

To circumvent the problems of embedding methods and provide theoretical support for network-response regression, we introduce a unifying intrinsic framework for network-response regression, by viewing each network as a random object \citep{mueller2016peter} lying in the space of graph Laplacians and adopting conditional \f means \citep{frec:48}. Specifically, let $G=(V, E)$ be a network with a set of nodes $V$ and a set of edge weights $E$. Under the assumption that $G$ is labeled and simple (i.e., there are no self-loops or multi edges), one can uniquely associate each network $G$ with its graph Laplacian $L$. Consider random pairs $(X, L)\sim F$,  where $X$ takes values in $\mathbb{R}^p$, $L$ is a graph Laplacian and $F$ indicates a suitable probability law. We investigate the dependence of $L$ on covariates of interest $X$ by adopting the general framework of \f regression \citep{mull:19:3, chen2020uniform}. A toy example is introduced in Figure~\ref{fig:exm} to illustrate the idea more comprehensively. The relationship between the observed networks and associated covariates in this toy example will be investigated using the proposed network-response regression.
The contributions of this paper are as follows: First, we provide  a precise characterization of the space of graph Laplacians, laying a foundation for the analysis of populations of networks represented as graph Laplacians,  where we adopt the power metric, with the Frobenius metric as a special case.  Second, we demonstrate that this characterization makes it possible to adopt \f regression for graph Laplacians. 
Third, the resulting network regression approach is shown to be competitive in finite sample situations when compared with previous network regression approaches \citep{severn2019manifold, severn2021non, zhang2018generalized}. Fourth, our methods are supported by theory, including pointwise and uniform rates of convergence. Fifth, we demonstrate the utility and flexibility of the proposed network-response regression model with fMRI data obtained from the ADNI neuroimaging study and also with New York taxi data.

The organization of this paper is as follows. In Section \ref{sec:prelim}, we provide a precise characterization of the space of graph Laplacians and discuss  metrics for this space. The proposed regression model for network responses and vector covariates is introduced in Section \ref{sec:model}. Pointwise and uniform rates of convergence for the estimators are established in Section \ref{sec:asym}. Computational details and simulation results for a sample of networks are presented in Section \ref{sec:sim}. The proposed framework is illustrated in Section \ref{sec:data} using the New York yellow taxi records and  rs-fMRI data from the ADNI study. Finally, we conclude with a brief discussion presented in Section \ref{sec:dis}. Detailed theoretical proofs and auxiliary results are in the Appendix. 

\section{Preliminaries}
\label{sec:prelim}
\subsection{Characterization of the Space of Graph Laplacians}
Let $G=(V, E)$ be a network with a set of nodes $V=\{v_1, \ldots, v_m\}$ and a set of edge weights $E=\{w_{ij}: w_{ij}\geq0, \, i, j=1, \ldots, m\}$, where $w_{ij}=0$ indicates $v_i$ and $v_j$ are unconnected. Some basic and mild restrictions on the networks $G$ we consider here are as follows.
\begin{enumerate}[label=(C\arabic*), leftmargin=1cm]
	\item $G$ is simple, i.e., there are no self-loops or multi-edges.\label{itm:a0}
	\item $G$ is weighted, undirected, and labeled.\label{itm:a1}
	\item The edge weights $w_{ij}$ are bounded above by $W\geq0$, i.e., $0\leq w_{ij}\leq W$.\label{itm:a2}
\end{enumerate}
Condition \ref{itm:a0} is required for the one-to-one correspondence between a network $G$ and its graph Laplacian, which is our central tool to represent networks. Condition \ref{itm:a1} guarantees that the adjacency matrix $A=(w_{ij})$ is symmetric, i.e., $w_{ij}=w_{ji}$ for all $i, j$. Condition \ref{itm:a2} puts a limit on the maximum strength of the connection between two nodes and prevents extremes. Any network satisfying Conditions \ref{itm:a0}--\ref{itm:a2} can be uniquely associated with its graph Laplacian $L=(l_{ij})$, 
\[l_{ij}=\begin{cases}-w_{ij},&i\neq j\\\sum_{k\neq i}w_{ik},&i=j\end{cases}\]
for $i, j=1, \ldots, m$, which motivates to characterize the space of networks through the corresponding space of graph Laplacians given by 
\begin{equation}
	\mathcal{L}_m=\{L=(l_{ij}):\,  L=L^\T; \, L1_m=0_m; \, -W\leq l_{ij}\leq0\; \, \text{for}\;i\neq j\},
	\label{eq:lmdef}
\end{equation}
where $1_m$ and $0_m$ are the $m$-vectors of ones and zeroes, respectively. Another well-known property of graph Laplacians is their positive semi-definiteness, $x^\T Lx\geq0\;\text{for all}\;x\in\mathbb{R}^m$, which immediately follows if $L\in\mathcal{L}_m$, as any such $L$ is diagonally dominant, i.e., $l_{ii}=\sum_{j\neq i}|l_{ij}|$ \citep[p.~232]{de:06}.

A precise geometric characterization of the space of graph Laplacians with fixed rank can be found in \citet{gine:17}. 
However, the fixed rank assumption is not practicable when considering network-response regression where the rank often will  change in dependence on predictor levels. 
This necessitates and motivates the study of the space of graph Laplacians without rank restrictions.
\begin{proposition}
	The space $\mathcal{L}_m$, defined in \eqref{eq:lmdef}, is a bounded, closed, and convex subset in $\mathbb{R}^{m^2}$ of dimension $m(m-1)/2$.
	\label{prop:lmmanifold}
\end{proposition}
All proofs are given in Appendix B. The convexity and closedness of $\mathcal{L}_m$ as shown in Proposition \ref{prop:lmmanifold} ensures the existence and uniqueness of projections onto $\mathcal{L}_m$ \citep[chap.~3]{deutsch2012best} that we will use in the proposed regression approach. 

\subsection{Choice of Metrics}
\label{subsec:metric}
One can choose between several metrics for the space of graph Laplacians $\mathcal{L}_m$. A common choice is the Frobenius metric, defined as
\begin{equation*}
	d_F(L_1, L_2)=\|L_1-L_2\|_F=\{\tr[(L_1-L_2)^\T(L_1-L_2)]\}^{1/2},
\end{equation*}
which corresponds to the usual Euclidean metric if the matrix is viewed as a vector of length $m^2$. 
While $d_F$ is the simplest of the possible metrics on the space of graph Laplacians, a downside of $d_F$ is the swelling effect, notably for positive semi-definite matrices \citep{arsigny2007geometric, lin2019riemannian}.

Denote the space of real symmetric positive semi-definite $m\times m$ matrices by $\mathcal{S}_m^+$. Note that the space of graph Laplacians $\mathcal{L}_m$ is a subset of $\mathcal{S}_m^+$. Let $U \Lambda U^\T$ be the usual spectral decomposition of $S\in\mathcal{S}_m^+$, with $U\in\mathcal{O}_m$ an orthogonal matrix and $\Lambda$ diagonal with nonnegative entries. Defining matrix power maps 
\begin{equation}
	\label{eq:falpha}
	F_{\alpha}(S)=S^\alpha=U\Lambda^\alpha U^\T: \mathcal{S}_m^+\mapsto\mathcal{S}_m^+,
\end{equation}
where the power $\alpha>0$ is a constant and noting  that $F_{\alpha}$ is bijective with inverse $F_{1/\alpha}$, the power metric \citep{dryd:09} between graph Laplacians is 
\begin{equation*}
	d_{F, \alpha}(L_1, L_2)=d_F[F_{\alpha}(L_1), F_{\alpha}(L_2)].
\end{equation*}

For $\alpha=1$, $d_{F, \alpha}$ reduces to the Frobenius metric $d_{F}$. For  larger $\alpha$, there is more emphasis  on larger entries of graph Laplacians, while for small $\alpha$, large and small entries are treated more evenly and there is less  sensitivity to outliers. In particular, $d_{F, \alpha}$ with $0<\alpha<1$ is associated with a reduced  swelling effect, while $d_{F, \alpha}$ with $\alpha>1$ in contrast will amplify it and thus often will be unfavorable. For $\alpha=1/2$ the square root metric $d_{F, 1/2}$ is a  canonical choice that has been widely studied \citep{dryd:09, dryden2010power, zhou:16, severn2019manifold, tavakoli2019spatial}. For example, \citet{dryden2010power} examined  different values  
of $\alpha$ in the context of  diffusion tensor imaging and ended up with the choice $\alpha=1/2$ and \citet{tavakoli2019spatial} also  demonstrated the advantages of using $d_{F, 1/2}$ when spatially modeling  linguistic object data. In our applications, we likewise focus on $d_{F, 1/2}$ due to its promising properties and compare its performance with the Frobenius metric $d_{F}$; see also \citet{petersen2016frechet} regarding  the choice of $\alpha$. 

\section{Network Regression}
\label{sec:model}
Consider a random object $Y\sim F_Y$ taking values in a metric space $(\Omega, d)$. Under appropriate conditions, the \f mean and \f variance of random objects in metric spaces \citep{frec:48}, as generalizations of usual notions of mean and variance, are defined as
\begin{equation*}
	\omega_{\oplus}=\argmin_{\omega\in\Omega}E[d^2(Y, \omega)],\quad V_{\oplus}=E[d^2(Y, \omega_{\oplus})],
\end{equation*}
where the existence and uniqueness of the minimizer depends on structural properties of the underlying metric space and is guaranteed for Hadamard spaces.
A general approach for regression of metric space-valued responses on Euclidean predictors is 
\f regression,  
for which both linear and locally linear versions have been developed \citep{mull:19:3, chen2020uniform}, and which we adopt as a tool to model regression relationships between responses which are $\mathcal{L}_m$-valued, i.e., graph Laplacians of fixed dimension $m$, and vectors of real-valued predictors. Equipped with a proper metric $d$, $\mathcal{L}_m$ becomes a metric space $(\mathcal{L}_m, d)$, and an analysis of the properties of this space 
is key to establish theory  
for the proposed network regression  models.  

Suppose $(X, L)\sim F$ is a random pair, where $X$ and $L$ take values in $\mathbb{R}^p$ and $\mathcal{L}_m \equiv (\mathcal{L}_m,d)$, respectively, and $F$ is the joint distribution of $(X, L)$ on $\mathbb{R}^p\times\mathcal{L}_m$. We denote the marginal distributions of $X$ and $L$ as $F_X$ and $F_L$, respectively, and assume that $\mu=E(X)$ and $\Sigma=\Var(X)$ exist, with $\Sigma$ positive definite. The conditional distributions $F_{X\mid L}$ and $F_{L\mid X}$ are also assumed to exist. The conditional \f mean, which corresponds to the regression function of $L$ given $X=x$, is 
\begin{equation}
	\label{eq:confm}
	m(x)=\argmin_{\omega\in\mathcal{L}_m}M(\omega, x),\quad M(\cdot, x)=E[d^2(L, \cdot)\mid X=x],
\end{equation}
where $M(\cdot, x)$ is referred to as the conditional \f function. Observing that the conventional conditional mean satisfies $E(Y|X=x)=\argmin_{y\in\mathbb{R}}E[(Y-y)^2|X=x]$ for real valued responses $Y$, the conditional \f mean can be viewed as a natural extension of the notion of a conditional mean  to network-valued and other metric space-valued responses,  where $(Y-y)^2$ is replaced by $d^2(L, \omega)$. Further suppose that $(X_k, L_k)\sim F, \, k=1, \ldots, n,$ are independent and define 
$$\bar{X}=n^{-1}\sum_{k=1}^nX_k, \quad \hat{\Sigma}=n^{-1}\sum_{k=1}^n(X_k-\bar{X})(X_k-\bar{X})^\T.$$

The global and local network regression are generalizations of multiple linear regression and local linear regression to network-valued responses. The key idea is to characterize the original regression functions as minimizers of weighted least squares problems and then to leverage the weights to define a weighted barycenter as an M-estimator, using the metric that is adopted for the  space of graph Laplacians.

The global network regression given $X=x$ is defined as
\begin{equation}
	\label{eq:gfm}
	m_{G}(x)=\argmin_{\omega\in\mathcal{L}_m}M_{G}(\omega, x),\quad M_{G}(\cdot, x)=E[s_{G}(x)d^2(L, \cdot)],
\end{equation}
where $s_{G}(x)=1+(X-\mu)^\T\Sigma^{-1}(x-\mu)$. The corresponding sample version is
\begin{equation}
	\label{eq:gfe}
	\hat{m}_{G}(x)=\argmin_{\omega\in\mathcal{L}_m}\hat{M}_{G}(\omega, x),\quad \hat{M}_{G}(\cdot, x)=\frac{1}{n}\sum_{k=1}^ns_{kG}(x)d^2(L_k, \cdot),
\end{equation}
where $s_{kG}(x)=1+(X_k-\bar{X})^\T\hat{\Sigma}^{-1}(x-\bar{X})$.

For local network regression, we present details only for the case of a scalar predictor $X\in\mathbb{R}$, where the extension to $X\in\mathbb{R}^p$ with $p>1$ is straightforward but may be subject to the curse of dimensionality. Consider a smoothing kernel $K(\cdot)$ corresponding to a one-dimensional probability density and $K_h(\cdot)=h^{-1}K(\cdot/h)$ with $h$ a bandwidth. The local network regression given $X=x$ is
\begin{equation}
\label{eq:lfm}
m_{L, h}(x)=\argmin_{\omega\in\mathcal{L}_m}M_{L, h}(\omega, x),\quad M_{L, h}(\cdot, x)=E[s_{L}(x, h)d^2(L, \cdot)],
\end{equation}
where $s_L(x, h)=K_h(X-x)[\mu_2-\mu_1(X-x)]/\sigma_0^2$ with $\mu_j=E[K_h(X-x)(X-x)^j]$ for $j=0, 1, 2$ and $\sigma_0^2=\mu_0\mu_2-\mu_1^2$. The corresponding sample version is
\begin{equation}
\label{eq:lfe}
\hat{m}_{L, n}(x)=\argmin_{\omega\in\mathcal{L}_m}\hat{M}_{L, n}(\omega, x),\quad \hat{M}_{L, n}(\cdot, x)=\frac{1}{n}\sum_{k=1}^ns_{kL}(x, h)d^2(L_k, \cdot).
\end{equation}
Here $s_{kL}(x, h)=K_h(X_k-x)[\hat{\mu}_2-\hat{\mu}_1(X_k-x)]/\hat{\sigma}_0^2$, where $\hat{\mu}_j=n^{-1}\sum_{k=1}^nK_h(X_k-x)(X_k-x)^j$ for $j=0, 1, 2$ and $\hat{\sigma}_0^2=\hat{\mu}_0\hat{\mu}_2-\hat{\mu}_1^2$. The dependence on $n$ is through the bandwidth sequence $h=h_n$.The local network regression estimate $\hat{m}_{L, n}(x)$, similar to global network regression, is an M-estimator that depends on the weights $s_{kL}(x, h)$.

For the case where $X\in\mathbb{R}^p$ with $p>1$, the weight function for local network regression takes a slightly different form,
\[s_L(x, h)=\frac{1}{\mu_0-\mu_1^\T\mu_2^{-1}\mu_1}K_h(X-x)[1-\mu_1^\T\mu_2^{-1}(X-x)],\]
where $\mu_0=E[K_h(X-x)], \mu_1=E[K_h(X-x)(X-x)]$, and $\mu_2=E[K_h(X-x)(X-x)(X-x)^\T]$ is nondegenerate. The sample version $s_{kL}(x, h)$ can be defined similarly.
While global network regression relies on stronger model assumptions, it does not require a tuning parameter and is applicable for categorical predictors. Local network regression, by contrast, is more flexible and may be preferable as long as the regression relation is smooth,  the covariate dimension is low and covariates are continuous.

\begin{figure}[tbp]
\centering
\includegraphics[width=0.45\linewidth]{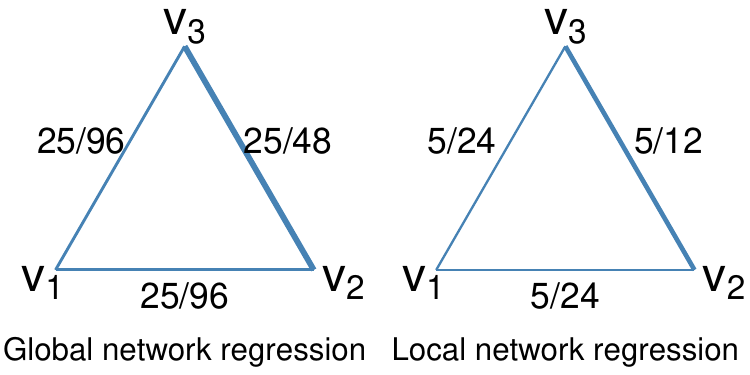}
\caption{Prediction networks at $X=5$ for the toy example as per Figure~\ref{fig:exm} using global and local network regression with the Frobenius metric $d_F$. The Epanechnikov kernel and a bandwidth $h=2$ are used in local network regression. The weight of each edge is marked beside the corresponding edge.}
\label{fig:exmpred}
\end{figure}

Consider the toy example in Figure~\ref{fig:exm}. In the case of global network regression, a simple calculation shows that the weight function is $s_{kG}(x)=1+(2k-5)(x-5)/5$ for $k=1, 2, 3, 4$. For local network regression, one can also obtain an explicit form of the weight function $s_{kL}(x, h)$ if a smoothing kernel $K(\cdot)$ and a bandwidth $h$ are specified. Endowed with the Frobenius metric $d_F$, the global and local network regression estimates as per \eqref{eq:gfe} and \eqref{eq:lfe} are 
\[\hat{m}_G(x)=\frac{1}{4}\sum_{k=1}^4s_{kG}(x)L_k,\quad \hat{m}_{L, n}(x)=\frac{1}{4}\sum_{k=1}^4s_{kL}(x, h)L_k.\]
Figure~\ref{fig:exmpred} shows the prediction networks at $X=5$ using global and local network regression, where the Epanechnikov kernel $K(u)=\frac{3}{4}(1-u^2)1_{[-1,1]}$ is used with a bandwidth $h=2$.
\section{Asymptotic Properties}
\label{sec:asym}
We  establish the consistency of both global and local network regression estimates as per \eqref{eq:gfe} and \eqref{eq:lfe},  using the metrics discussed in Section \ref{subsec:metric}. Both pointwise and uniform rates of convergence are obtained under the framework of M-estimation. For both Frobenius metric $d_F$ and power metric $d_{F, \alpha}$ with $0<\alpha<1$, the pointwise rates of convergence are optimal for both global and local network regression in the sense that as generalizations of multiple linear regression and local linear regression for the case of Euclidean responses,  they correspond to the known optimal rates for the Euclidean special case. These rates are validated by simulations in Section \ref{subsec:simurates}, where we find that the empirical convergence when sample size is increasing   are entirely consistent with the theoretical predictions. We first consider the case where $\mathcal{L}_m$ is endowed with the Frobenius metric $d_F$. 
The convexity and closedness of $\mathcal{L}_m$ implies that  the minimizers $\hat{m}_G(x)$ and $\hat{m}_{L, n}(x)$ defined in \eqref{eq:gfe} and \eqref{eq:lfe} exist and are unique for any $x$. The following result formalizes the consistency of the  proposed global network regression estimate and provides rates of convergence, where $\|\cdot\|_E$ denotes the Euclidean norm in $\mathbb{R}^p$.

\begin{theorem}
\label{thm:euclidean:g}
Adopting the space of graph Laplacians $\mathcal{L}_m$ endowed with the Frobenius metric $d_F$, for a fixed $x\in\mathbb{R}^p$, it holds that for $m_{G}(x)$ and $\hat{m}_G(x)$ as per \eqref{eq:gfm} and \eqref{eq:gfe} that
\begin{equation}
	\label{eq:rcge}
	d_F[m_{G}(x), \hat{m}_{G}(x)]=O_p(n^{-1/2}).
\end{equation}
Furthermore, for a given $B>0$ and any given $\varepsilon>0$,
\begin{equation*}
	\sup_{\|x\|_E\leq B}d_F[m_{G}(x), \hat{m}_{G}(x)]=O_p(n^{-1/[2(1+\varepsilon)]}).
\end{equation*}
\end{theorem}

The pointwise rate of convergence for conventional multiple linear regression is well known to be $O_p(n^{-1/2})$. The pointwise rate of convergence in Theorem \ref{thm:euclidean:g} is thus optimal in the sense that it remains the same as the optimal rate of multiple linear regression. Analogous to the Euclidean case, with its well-known bias-variance trade-off for nonparametric regression, the rate of convergence for the local network regression estimator depends both on the  metric equivalent of bias, which is  $d_F[m(x), m_{L, h}(x)]$,  as well as on the  stochastic deviation $d_F[m_{L, h}(x), \hat{m}_{L, n}(x)]$; see the following result. The kernel and distributional assumptions \ref{itm:lp0}--\ref{itm:lu1} in the Appendix are standard for local regression estimation.

\begin{theorem}
\label{thm:euclidean:l}
If the space of graph Laplacians $\mathcal{L}_m$ is endowed with the Frobenius metric $d_F$ and Assumptions \ref{itm:lp0}, \ref{itm:lp1} hold, then for a fixed $x\in\mathbb{R}$ and   $m(x), m_{L, h}(x)$, and $\hat{m}_{L, n}(x)$ as per \eqref{eq:confm}, \eqref{eq:lfm}, and \eqref{eq:lfe},
\begin{equation*}
	d_F[m(x), m_{L, h}(x)]=O(h^{2}),
\end{equation*}
\begin{equation*}
	d_F[m_{L, h}(x), \hat{m}_{L, n}(x)]=O_p[(nh)^{-1/2}], 
\end{equation*}
and with $h\sim n^{-1/5}$,
\begin{equation}
	\label{eq:rcle}
	d_F[m(x), \hat{m}_{L, n}(x)]=O_p(n^{-2/5}).
\end{equation}
Under Assumptions \ref{itm:lu0}, \ref{itm:lu1} for a  closed interval $\mathcal{T}\subset\mathbb{R}$, if $h\to0$, $nh^2(-\log h)^{-1}\to\infty$ as $n\to\infty$, then for any $\varepsilon>0$,
\begin{equation*}
	\sup_{x\in\mathcal{T}}d_F[m(x), m_{L, h}(x)]=O(h^{2}),
\end{equation*}
\begin{equation*}
	\sup_{x\in\mathcal{T}}d_F[m_{L, h}(x), \hat{m}_{L, n}(x)]=O_p(\max\{(nh^2)^{-1/(2+\varepsilon)}, [nh^2(-\log h)^{-1}]^{-1/2}\}),
\end{equation*}
and with $h\sim n^{-1/(6+2\varepsilon)}$, 
\begin{equation*}
	\sup_{x\in\mathcal{T}}d_F[m(x), \hat{m}_{L, n}(x)]=O_p(n^{-1/(3+\varepsilon)}).
\end{equation*}
\end{theorem}

The proofs of these results rely on the fact that $\mathcal{L}_m$ is convex, bounded, and of finite dimension. We represent  global and local network regressions as projections $P_{\mathcal{L}_m}$ onto $\mathcal{L}_m$,
\begin{align*}
m_{G}(x)=\argmin_{\omega\in\mathcal{L}_m}d_F^2[B_{G}(x), \omega]=P_{\mathcal{L}_m}[B_{G}(x)],\\
m_{L, h}(x)=\argmin_{\omega\in\mathcal{L}_m}d_F^2[B_{L, h}(x), \omega]=P_{\mathcal{L}_m}[B_{L, h}(x)],
\end{align*}
where $B_{G}(x)=E[s_{G}(x)L]$ and $B_{L, h}(x)=E[s_{L}(x, h)L]$. 
The corresponding sample versions are
\begin{align*}
\hat{m}_{G}(x)=\argmin_{\omega\in\mathcal{L}_m}d_F^2[\hat{B}_{G}(x), \omega]=P_{\mathcal{L}_m}[\hat{B}_{G}(x)],\\
\hat{m}_{L, n}(x)=\argmin_{\omega\in\mathcal{L}_m}d_F^2[\hat{B}_{L, n}(x), \omega]=P_{\mathcal{L}_m}[\hat{B}_{L, n}(x)],
\end{align*}
where $\hat{B}_{G}(x)=n^{-1}\sum_{k=1}^ns_{kG}(x)L_k$ and $\hat{B}_{L, n}(x)=n^{-1}\sum_{k=1}^ns_{kL}(x, h)L_k$.

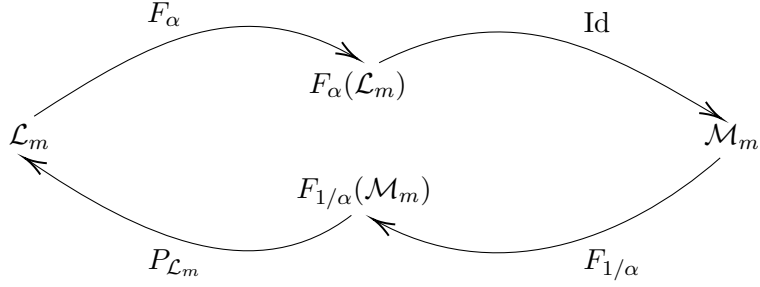
\begin{figure}[t]
\centering      
\begin{tikzpicture}[x=0.75pt,y=0.75pt,yscale=-1,xscale=1]
	\draw    (104,109) .. controls (163.7,70.19) and (208.75,46.01) .. (266.13,79.49) ;
	\draw [shift={(267,80)}, rotate = 210.64] [color={rgb, 255:red, 0; green, 0; blue, 0 }  ][line width=0.75]    (10.93,-3.29) .. controls (6.95,-1.4) and (3.31,-0.3) .. (0,0) .. controls (3.31,0.3) and (6.95,1.4) .. (10.93,3.29)   ;
	\draw    (278,82) .. controls (345.66,47.94) and (392.22,75.64) .. (449.63,109.64) ;
	\draw [shift={(450.49,110.15)}, rotate = 210.64] [color={rgb, 255:red, 0; green, 0; blue, 0 }  ][line width=0.75]    (10.93,-3.29) .. controls (6.95,-1.4) and (3.31,-0.3) .. (0,0) .. controls (3.31,0.3) and (6.95,1.4) .. (10.93,3.29)   ;
	\draw    (450,130) .. controls (400.43,173.59) and (335.16,195.02) .. (275.89,161.51) ;
	\draw [shift={(275,161)}, rotate = 389.91999999999996] [color={rgb, 255:red, 0; green, 0; blue, 0 }  ][line width=0.75]    (10.93,-3.29) .. controls (6.95,-1.4) and (3.31,-0.3) .. (0,0) .. controls (3.31,0.3) and (6.95,1.4) .. (10.93,3.29)   ;
	\draw    (264,159) .. controls (216.64,195.34) and (159.36,171.58) .. (100.18,130.62) ;
	\draw [shift={(99.28,130.11)}, rotate = 389.91999999999996] [color={rgb, 255:red, 0; green, 0; blue, 0 }  ][line width=0.75]    (10.93,-3.29) .. controls (6.95,-1.4) and (3.31,-0.3) .. (0,0) .. controls (3.31,0.3) and (6.95,1.4) .. (10.93,3.29)   ;
	\draw (90.03,112) node [anchor=north west][inner sep=0.75pt]    {$\mathcal{L}_{m}$};
	\draw (241.29,85) node [anchor=north west][inner sep=0.75pt]    {$F_{\alpha}(\mathcal{L}_{m})$};
	\draw (440.64,112) node [anchor=north west][inner sep=0.75pt]    {$\mathcal{M}_{m}$};
	\draw (159.69,50) node [anchor=north west][inner sep=0.75pt]    {$F_{\alpha}$};
	\draw (379.79,175) node [anchor=north west][inner sep=0.75pt]    {$F_{1/\alpha }$};
	\draw (235,138) node [anchor=north west][inner sep=0.75pt]    {$F_{1/\alpha}(\mathcal{M}_{m})$};
	\draw (379.99,55) node [anchor=north west][inner sep=0.75pt]    {$\mathrm{Id}$};
	\draw (160,175) node [anchor=north west][inner sep=0.75pt]    {$P_{\mathcal{L}_{m}}$};
\end{tikzpicture}
\caption{Schematic diagram for the power metric $d_{F, \alpha}$, where $\mathcal{L}_m$ and $\mathcal{M}_m$ are the space of graph Laplacians and the embedding space defined in \eqref{eq:lmdef} and \eqref{eq:mm}, respectively. Network regression is carried out using the Frobenius metric $d_F$ in the embedding space $\mathcal{M}_m$. Here  $F_{\alpha}$ is the matrix power map defined in \eqref{eq:falpha} and  $F_{1/\alpha}$  its inverse, while  $F_{\alpha}(\mathcal{L}_m)$ is the image of $\mathcal{L}_m$ under $F_{\alpha}$ and $F_{1/\alpha}(\mathcal{M}_m)$ is the image of $\mathcal{M}_m$ under $F_{1/\alpha}$. The identity map  $\mathrm{Id}$  embeds $F_{\alpha}(\mathcal{L}_m)$ into $\mathcal{M}_m$, where  $P_{\mathcal{L}_m}$ is the projection onto $\mathcal{L}_m$.}
\label{fig:diagramalpha}
\end{figure}

Next considering the power metric $d_{F, \alpha}$, recall that  the graph Laplacian $L$ is centered and the off-diagonal entries are bounded by $W$ as per \eqref{eq:lmdef}. By the equivalence of the Frobenius norm and the $l_2$-operator norm in $\mathbb{R}^{m^2}$, it immediately follows that the largest eigenvalue of $L$ is bounded, say by $D$, a nonnegative constant depending on $m$ and $W$. Define the embedding space $\mathcal{M}_m$ as 
\begin{equation}
\label{eq:mm}
\mathcal{M}_m=\{S\in\mathcal{S}_m^+: \lambda_1(S)\leq D^\alpha\},
\end{equation}
where $\lambda_1(S)$ is the largest eigenvalue of $S$. The image of $\mathcal{L}_m$ under the matrix power map $F_\alpha$, i.e., $F_{\alpha}(\mathcal{L}_m)$, is a subset of $\mathcal{M}_m$ as a consequence of the bound $D$ on the largest eigenvalue of each graph Laplacian. 
After applying the matrix power map $F_\alpha$, the image of $\mathcal{L}_m$ is  embedded in $\mathcal{M}_m$, where network regression is carried out using the Frobenius metric $d_F$. When transforming back from the embedding space $\mathcal{M}_m$ to $\mathcal{L}_m$, we first apply the inverse matrix power map $F_{1/\alpha}$ and then a projection $P_{\mathcal{L}_m}$ onto $\mathcal{L}_m$. The general idea involving embedding, mapping and projections is shown schematically in Figure~\ref{fig:diagramalpha}.
The global network regression in the embedding space $\mathcal{M}_m$ using the Frobenius metric $d_F$ also uses the projection $P_{\mathcal{M}_m}$ onto $\mathcal{M}_m$,
\begin{equation}
\label{eq:gfm:alpha:m}
m_{G}^\alpha(x)=\argmin_{\omega\in\mathcal{M}_m}d_F^2[B_{G}^\alpha(x), \omega]=P_{\mathcal{M}_m}[B_{G}^\alpha(x)],
\end{equation}
where $B_{G}^\alpha(x)=E[s_{G}(x)F_{\alpha}(L)]$. Then the global network regression in the space of graph Laplacians $\mathcal{L}_m$ using the power metric $d_{F, \alpha}$ is obtained by applying the inverse matrix power map $F_{1/\alpha}$ and projection $P_{\mathcal{L}_m}$ successively to $m_{G}^\alpha(x)$, i.e., 
\begin{equation}
\label{eq:gfm:alpha}
m_{G}(x)=\argmin_{\omega\in\mathcal{L}_m}d_F^2\{F_{1/\alpha}[m_{G}^\alpha(x)], \omega\}=P_{\mathcal{L}_m}\{F_{1/\alpha}[m_{G}^\alpha(x)]\}.
\end{equation}
The corresponding sample versions are
\begin{align}
\hat{m}_{G}^\alpha(x)=\argmin_{\omega\in\mathcal{M}_m}d_F^2[\hat{B}_{G}^\alpha(x), \omega]=P_{\mathcal{M}_m}[\hat{B}_{G}^\alpha(x)],\nonumber\\
\hat{m}_{G}(x)=\argmin_{\omega\in\mathcal{L}_m}d_F^2\{F_{1/\alpha}[\hat{m}_{G}^\alpha(x)], \omega\}=P_{\mathcal{L}_m}\{F_{1/\alpha}[\hat{m}_{G}^\alpha(x)]\},\label{eq:gfe:alpha}
\end{align}
where $\hat{B}_{G}^\alpha(x)=n^{-1}\sum_{k=1}^ns_{kG}(x)F_{\alpha}(L_k)$. Similarly for the local network regression,
\begin{align}
m(x)=P_{\mathcal{L}_m}(F_{1/\alpha}\{P_{\mathcal{M}_m}[B^\alpha(x)]\}),\label{eq:confm:alpha}\\
m_{L, h}(x)=P_{\mathcal{L}_m}(F_{1/\alpha}\{P_{\mathcal{M}_m}[B_{L, h}^\alpha(x)]\}),\label{eq:lfm:alpha}\\
\hat{m}_{L, n}(x)=P_{\mathcal{L}_m}(F_{1/\alpha}\{P_{\mathcal{M}_m}[\hat{B}_{L, n}^\alpha(x)]\}),\label{eq:lfe:alpha}
\end{align}
where $\quad B^\alpha(x)=E[F_{\alpha}(L)\mid X=x]$, $\quad B_{L, h}^\alpha(x)=E[s_{L}(x, h)F_{\alpha}(L)]\quad$ and \newline $\hat{B}_{L, n}^\alpha(x)=n^{-1}\sum_{k=1}^ns_{kL}(x, h)F_{\alpha}(L_k)$.

To obtain rates of convergence for power metrics $d_{F, \alpha}$, an auxiliary result 
on the H\"{o}lder continuity of the matrix power map $F_{\alpha}$ is needed.  For $U$ a set in $\mathbb{R}^{n_1}$, $E$  a nonempty subset of $U$ and $0<\beta\leq1$, a function $g: U\mapsto\mathbb{R}^{n_2}$ is uniformly H\"{o}lder continuous with order  $\beta$ and coefficient $H$ on $E$, i.e., $(\beta, H)$-H\"{o}lder continuous, if there exists $H\geq0$ such that
\begin{equation*}
\|g(x)-g(y)\|_F\leq H\|x-y\|_F^{\beta},\quad\text{for all}\;x, y\in E.
\end{equation*}
For $\beta=1$ the function $g$ is said to be $H$-Lipschitz continuous on $E$.
\begin{proposition}
\label{prop:falpha}
For $\mathcal{E}_m=\{S\in\mathcal{S}_m^+: \lambda_1(S)\leq C\}$, where $\lambda_1(S)$ is the largest eigenvalue of $S$ and $C\geq0$ is a constant,  the matrix power map $F_{\alpha}$ as per \eqref{eq:falpha} is
\begin{enumerate}[label=(\roman*)]
	\item $(\alpha, m^{(1-\alpha)/2})$-H\"{o}lder continuous on $\mathcal{S}_m^+$\,\;for $0<\alpha<1$; \label{itm:holder}
	\item $\alpha C^{\alpha-1}$-Lipschitz continuous on $\mathcal{E}_m$\,\;for $\alpha\geq1$.\label{itm:lipschitz}
\end{enumerate}
\end{proposition}
Proposition \ref{prop:falpha} leads to rate of convergence results for the global and local network regression estimators defined in \eqref{eq:gfe:alpha} and \eqref{eq:lfe:alpha}, where the population targets for the global and local network regression under the power metric $d_{F, \alpha}$ are defined in \eqref{eq:gfm:alpha} and \eqref{eq:confm:alpha}.

\begin{theorem}
\label{thm:euclideanpower:g}
If the space of graph Laplacians $\mathcal{L}_m$ is endowed with the power metric $d_{F, \alpha}$, for any fixed $x\in\mathbb{R}^p$, it holds for $m_{G}(x)$ and $\hat{m}_G(x)$ as per \eqref{eq:gfm:alpha} and \eqref{eq:gfe:alpha} that
\begin{equation}
	\label{eq:rcgp}
	d_F[m_{G}(x), \hat{m}_{G}(x)]=\begin{cases}O_p(n^{-1/2})&0<\alpha\leq1\\O_p(n^{-1/(2\alpha)})&\alpha>1\end{cases}.
\end{equation}
Furthermore, for any $B>0$ and 	 any $\varepsilon>0$
\begin{equation*}
	\sup_{\|x\|_E\leq B}d_F[m_{G}(x), \hat{m}_{G}(x)]=\begin{cases}O_p(n^{-1/[2(1+\varepsilon)]})&0<\alpha\leq1\\O_p(n^{-1/[2(1+\varepsilon)\alpha]})&\alpha>1\end{cases}.
\end{equation*}
\end{theorem}

\begin{theorem}
\label{thm:euclideanpower:l}
Suppose the space of graph Laplacians $\mathcal{L}_m$ is endowed with the power metric $d_{F, \alpha}$. Under Assumptions \ref{itm:lp0}, \ref{itm:lp1},  for a fixed $x\in\mathbb{R}$, it holds for $m(x), m_{L, h}(x)$, and $\hat{m}_{L, n}(x)$ as per \eqref{eq:confm:alpha}, \eqref{eq:lfm:alpha}, and \eqref{eq:lfe:alpha}, respectively,  that
\begin{equation*}
	d_F[m(x), m_{L, h}(x)]=\begin{cases}O(h^{2})&0<\alpha\leq1\\O(h^{2/\alpha})&\alpha>1\end{cases},
\end{equation*}
\begin{equation*}
	d_F[m_{L, h}(x), \hat{m}_{L, n}(x)]=\begin{cases}O_p[(nh)^{-1/2}]&0<\alpha\leq1\\O_p[(nh)^{-1/(2\alpha)}]&\alpha>1\end{cases},
\end{equation*}
and with $h\sim n^{-1/5}$, 
\begin{equation}
	\label{eq:rclp}
	d_F[m(x), \hat{m}_{L, n}(x)]=\begin{cases}O_p(n^{-2/5})&0<\alpha\leq1\\O_p(n^{-2/(5\alpha)})&\alpha>1\end{cases}.
\end{equation}
If Assumptions \ref{itm:lu0}, \ref{itm:lu1} hold for a given closed interval $\mathcal{T}\subset\mathbb{R}$ and  $h\to0$, $nh^2(-\log h)^{-1}\to\infty$ as $n\to\infty$, then for any $\varepsilon>0$,
\begin{equation*}
	\sup_{x\in\mathcal{T}}d_F[m(x), m_{L, h}(x)]=\begin{cases}O(h^{2})&0<\alpha\leq1\\O(h^{2/\alpha})&\alpha>1\end{cases},
\end{equation*}
\begin{align*}
	&\sup_{x\in\mathcal{T}}d_F[m_{L, h}(x), \hat{m}_{L, n}(x)]=
	\nonumber\\&
	\begin{cases}O_p(\max\{(nh^2)^{-1/(2+\varepsilon)}, [nh^2(-\log h)^{-1}]^{-1/2}\})&0<\alpha\leq1\\O_p(\max\{(nh^2)^{-1/[(2+\varepsilon)\alpha]}, [nh^2(-\log h)^{-1}]^{-1/(2\alpha)}\})&\alpha>1\end{cases},
\end{align*}
and with $h\sim n^{-1/(6+2\varepsilon)}$, 
\begin{equation*}
	\sup_{x\in\mathcal{T}}d_F[m(x), \hat{m}_{L, n}(x)]=\begin{cases}O_p(n^{-1/(3+\varepsilon)})&0<\alpha\leq1\\O_p(n^{-1/[(3+\varepsilon)\alpha]})&\alpha>1\end{cases}.
\end{equation*}
\end{theorem}

For the power metric $d_{F, \alpha}$, rates of convergence for both the global and local network regression depend on the choice of power $\alpha$. Specifically, rates of convergence are the same as those for the Frobenius metric $d_{F}$ if $0<\alpha<1$. When $\alpha>1$, we observe that rates of convergence are sub-optimal as $\alpha$ appears as a denominator. Therefore, the power metric $d_{F, \alpha}$ with $\alpha>1$ is generally not recommended, while whether or not to use power metric with $0<\alpha<1$ is not affecting the convergence and thus remains a matter of interpretation and other application-specific considerations.
The convexity of the target space is crucial in the proof of existence and uniqueness for the minimizers in \eqref{eq:confm}--\eqref{eq:lfe}. 
Since uniqueness for the minimizers in \eqref{eq:confm}--\eqref{eq:lfe} cannot be guaranteed,  we include the  embedding $F_{\alpha}(\mathcal{L}_m)$ in $\mathcal{M}_m$ as defined in \eqref{eq:mm}, which ensures  uniqueness for the minimizers in \eqref{eq:confm}--\eqref{eq:lfe}. 

\section{Implementation and Simulations}
\label{sec:sim}
\subsection{Implementation Details}
Implementation of the proposed method involves two projections $P_{\mathcal{L}_m}$ and $P_{\mathcal{M}_m}$ as mentioned in Section \ref{sec:asym}. Due to the convexity and closedness of $\mathcal{L}_m$ and $\mathcal{M}_m$, $P_{\mathcal{L}_m}$ and $P_{\mathcal{M}_m}$ exist and are unique. 
To implement $P_{\mathcal{L}_m}(B)$ where $B=(b_{ij})$ is a constant $m\times m$ matrix, one needs to solve 
\begin{align}
\begin{split}
	&\text{minimize}\quad f(L)=d^2_F(B, L)=\sum_{i=1}^m\sum_{j=1}^m(b_{ij}-l_{ij})^2\\
	&\text{subject to}\quad l_{ij}-l_{ji}=0,\quad i, j=1, \ldots, m,\\
	&\hspace{0.93in}\sum_{j=1}^ml_{ij}=0,\quad i=1, \ldots, m,\\
	&\hspace{0.63in}-W\leq l_{ij}\leq0,\quad i, j=1, \ldots, m; \, i\neq j,
\end{split}
\label{eq:convex}
\end{align}
where $L=(l_{ij})$ is a graph Laplacian. The objective function $f(L)$ is convex quadratic since its Hessian $2I_{m^2}$ is strictly positive definite. The three constraints, corresponding to the definition of $\mathcal{L}_m$ in \eqref{eq:lmdef}, are all linear so that \eqref{eq:convex} is a convex quadratic optimization problem, for the practical solution of which 
we use the \texttt{osqp} \citep{osqp} package in R \citep{rcoreteam}.

Note that $\mathcal{M}_m$ is a bounded subset of the positive semi-definite cone $\mathcal{S}_m^+$. To implement $P_{\mathcal{M}_m}$, we first  project on $\mathcal{S}_m^+$ and then truncate the eigenvalues to ensure that the largest eigenvalue is less than or equal to $D^\alpha$. The projection $P_{\mathcal{S}_m^+}$ on $\mathcal{S}_m^+$ is straightforward and has been studied in \citet[p.~399]{boyd:04}. The unique solution for $P_{\mathcal{S}_m^+}(B)$ is $\sum_{i=1}^m\max\{0, \lambda_i\}v_iv_i^\T$, where $B=\sum_{i=1}^m\lambda_iv_iv_i^\T$ is the spectral decomposition of a constant $m\times m$ symmetric matrix $B$. 

\begin{figure}[tbp]
\centering
\includegraphics[width=0.9\linewidth]{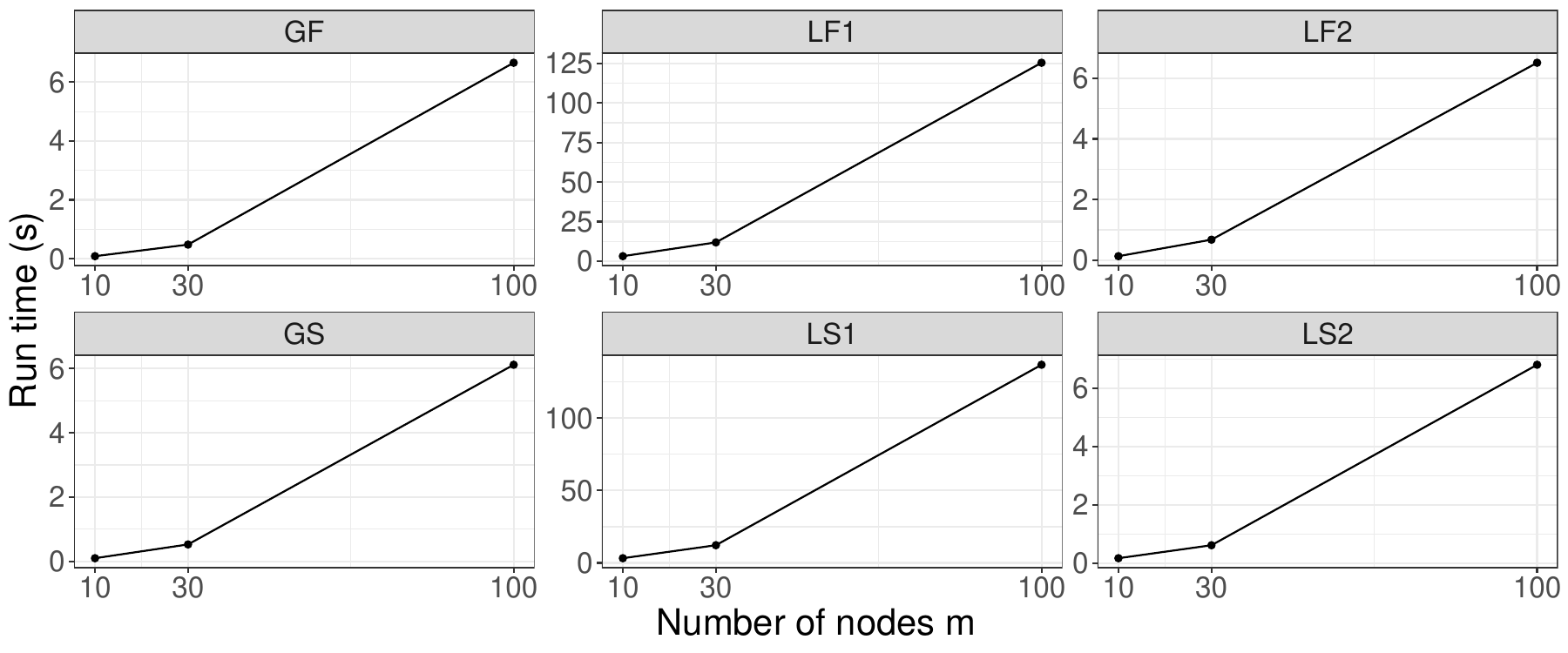}
\caption{Run time of the proposed regression models in seconds for different number of nodes $m$. GF, global network regression using the Frobenius metric; GS, global network regression using the square root metric; LF1, local network regression using the Frobenius metric with no pre-specified bandwidth; LS1, local network regression using the square root metric with no pre-specified bandwidth; LF2, local network regression using the Frobenius metric with pre-specified bandwidth; LS2, local network regression using the square root metric with pre-specified bandwidth.}
\label{fig:rt}
\end{figure}

R codes for the proposed regression models and numerical simulations are available at \url{https://github.com/yidongzhou/Network-Regression-with-Graph-Laplacians}. The run time of the proposed regression models for different number of nodes $m$ using R version 4.2.0 (2022-04-22) running under Darwin on MacBook Pro M1 are summarized in Figure~\ref{fig:rt}.

\subsection{Simulation Assessing Rates of Convergence}
\label{subsec:simurates}
To assess the performance of the global and local network regression estimates in \eqref{eq:gfe} and \eqref{eq:lfe} through simulations, we need to devise a generative model. 
Denote the half vectorization excluding the diagonal of a symmetric and centered matrix by $\mathrm{vech}$, with inverse operation  $\mathrm{vech}^{-1}$. By the symmetry and centrality as per \eqref{eq:lmdef}, every graph Laplacian $L$ is fully determined by its upper (or lower) triangular part, which can be further vectorized into $\mathrm{vech}(L)$, a vector of length $d=m(m-1)/2$. We construct the conditional distributions $F_{L\mid X}$ by assigning an independent beta distribution to each element of $\mathrm{vech}(L)$. Specifically, a random sample $(\beta_1, \ldots, \beta_d)^\T$ is generated using beta distributions whose parameters depend on the scalar predictor $X$ and vary under different simulation scenarios. The random response $L$ is then generated conditional on $X$ through an inverse half vectorization $\mathrm{vech}^{-1}$ applied to  $(-\beta_1, \ldots, -\beta_d)^\T$. We included four different simulation scenarios involving different types of regression and  metrics, which  are summarized  in Table~\ref{tab:simusetting}. 
\begin{table}[tb]
\centering
\begin{tabular}{l|lll}
	\hline
	& Type   & Metric       & Setting                                                     \\\hline
	\shortstack[l]{I\\\vspace{2em}}       & \shortstack[l]{\vspace{1em}\\global\\network regression\\\vspace{1.5em}}& \shortstack[l]{$d_F$\\\vspace{2em}}        & \shortstack[l]{\\$m(x)=\mathrm{vech}^{-1}(-x, \ldots, -x)$,\\$L=\mathrm{vech}^{-1}(-\beta_1, \ldots, -\beta_d)$, \\where $\beta_j\overset{i.i.d.}{\sim}\mathrm{Beta}(X, 1-X)$\\\vspace{0.2em}}                                          \\ 
	\shortstack[l]{II\\\vspace{2em}}      & \shortstack[l]{\vspace{1em}\\global\\network regression\\\vspace{1.5em}}& \shortstack[l]{$d_{F, 1/2}$\\\vspace{2em}} & \shortstack[l]{
		\\$m(x)=P_{\mathcal{L}_m}(F_2\{P_{\mathcal{M}_m}[\mathrm{vech}^{-1}(-x, \ldots, -x)]\})$,\\$L=F_2[\mathrm{vech}^{-1}(-\beta_1, \ldots, -\beta_d)]$, \\where $\beta_j\overset{i.i.d.}{\sim}\mathrm{Beta}(X, 1-X)$\\\vspace{0.2em}}                                     \\ 
	\shortstack[l]{III\\\vspace{2em}}     & \shortstack[l]{\vspace{1em}\\local\\network regression\\\vspace{1.5em}}& \shortstack[l]{$d_F$\\\vspace{2em}} & \shortstack[l]{\\$m(x)=\mathrm{vech}^{-1}[-\mathrm{sin}(\pi x), \ldots, -\mathrm{sin}(\pi x)]$,\\$L=\mathrm{vech}^{-1}(-\beta_1, \ldots, -\beta_d)$, \\where $\beta_j\overset{i.i.d.}{\sim}\mathrm{Beta}[\mathrm{sin}(\pi X), 1-\mathrm{sin}(\pi X)]$\\\vspace{0.2em}}      \\ 
	\shortstack[l]{IV\\\vspace{1.3em}}     & \shortstack[l]{\vspace{1em}\\local\\network regression\\\vspace{0.8em}}& \shortstack[l]{$d_{F, 1/2}$\\\vspace{1.3em}} & \shortstack[l]{\\$m(x)=	P_{\mathcal{L}_m}[F_2(P_{\mathcal{M}_m}\{E[F_{1/2}(L)\mid X=x]\})]$,\\$L=\mathrm{vech}^{-1}(-\beta_1, \ldots, -\beta_d)$, \\where $\beta_j\overset{i.i.d.}{\sim}\mathrm{Beta}[\mathrm{sin}(\pi X), 1-\mathrm{sin}(\pi X)]$}\\\hline
\end{tabular}
\caption{Four different simulation scenarios,  where $m$ is the true regression function and  $L$ represents the generated random response. 
	The parameters for the beta distributions of the random variables  $\beta_j$ depend on the predictors  $X$ as indicated for simulation scenarios I - IV.}
\label{tab:simusetting}
\end{table}
The space of graph Laplacians $\mathcal{L}_m$ is not a vector space. Instead, it is a bounded, closed, and convex subset in $\mathbb{R}^{m^2}$ of dimension $m(m-1)/2$ as shown in Proposition \ref{prop:lmmanifold}. To ensure that the random response $L$ generated in simulations resides in $\mathcal{L}_m$, the off-diagonal entries $-\beta_i, i=1, \ldots, d,$ need to be nonpositive and bounded below as per \eqref{eq:lmdef}. To this end,  $\beta_i, i=1, \ldots, d$ are sampled from beta distributions, which are defined on the interval $(0, 1)$ and whose parameters depend on the uniformly distributed predictor $X$.

The consistency of global network regression relies on the assumption that the true regression function $m(x)$ is equal to $m_G(x)$ as defined in \eqref{eq:confm} and \eqref{eq:gfm}, respectively. This assumption is satisfied if each entry of the graph Laplacian $L$ is conditionally linear in the predictor $X$. For the global network regression under the square root metric $d_{F, 1/2}$, an extra matrix square map $F_2$ is required to ensure that the global network regression estimate in the metric space $(\mathcal{M}_m, d_F)$ as per \eqref{eq:gfm:alpha:m} with $\alpha=1/2$ is element-wise linear in $X$.

\begin{figure}[t]
\centering
\includegraphics[width=0.9\linewidth]{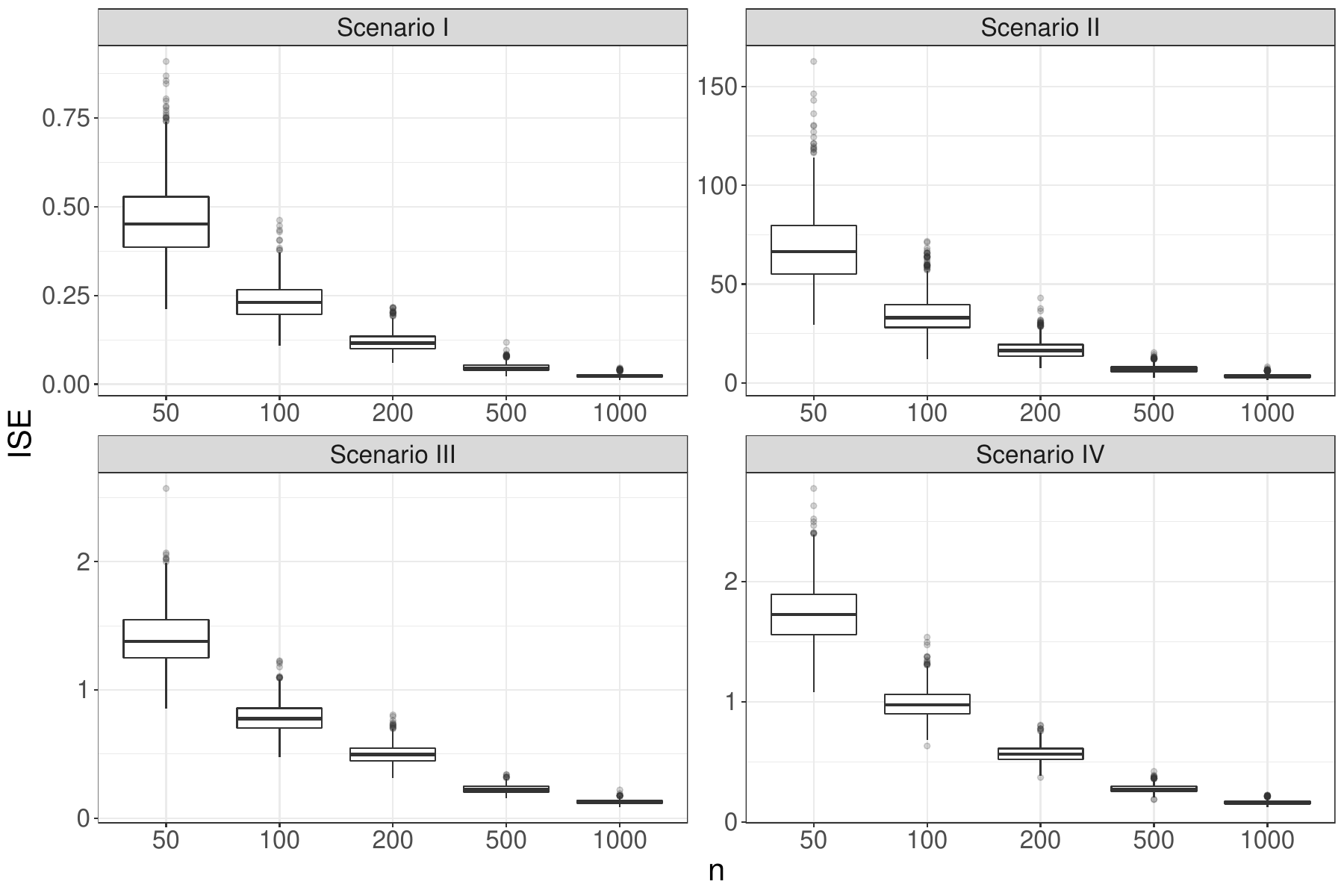}
\caption{Boxplots of integrated square errors (ISE) for five sample sizes under four simulation scenarios.}
\label{fig:simu}
\end{figure}

We investigated sample sizes $n=50, 100, 200, 500, 1000$, with $Q=1000$ Monte Carlo runs for each simulation scenario. In each iteration, random samples of pairs $(X_k, L_k)$, $k=1, \ldots, n$ were generated by sampling $X_k\sim\mathrm{U}(0, 1)$, setting $m=10$, and following the above procedure. For the $q$th simulation of a particular sample size, with $\hat{m}_q(x)$ denoting the fitted regression function, the quality of the estimation was quantified  by the integrated squared error $\mathrm{ISE}_q=\int_0^1d_F^2[m(x), \hat{m}_q(x)]dx$, where $m(x)$ is the true regression function and the  average quality of the estimation over the $Q=1000$ Monte Carlo runs was assessed  by the mean integrated squared error
\begin{equation}
\label{eq:mise}
\mathrm{MISE}=\frac{1}{Q}\sum_{q=1}^Q\int_0^1d_F^2[m(x), \hat{m}_q(x)]dx.
\end{equation}
The bandwidths for the local network regression in simulation scenarios III and IV were chosen by  leave-one-out cross-validation. The results are summarized in Figure~\ref{fig:simu} and Table~\ref{tab:simuresults}.
\begin{table}[t]
\centering
\begin{tabular}{l|cccc}
\hline
Sample size & I & II & III & IV\\\hline
50 & 0.465 & 68.63 & 1.404 & 1.738\\
100 & 0.235 & 34.36 & 0.786 & 0.983\\
200 & 0.119 & 16.82 & 0.502 & 0.566\\
500 & 0.047 & 6.87 & 0.226 & 0.275\\
1000 & 0.024 & 3.44 & 0.127 & 0.163\\\hline
\end{tabular}
\caption{Mean integrated squared errors (MISE) for five samples sizes under four simulation scenarios.}
\label{tab:simuresults}
\end{table}
With increasing sample size, ISE is seen to decrease for all scenarios, 
demonstrating the convergence  of network regression to the target. The empirical rate of convergence under each simulation scenario was assessed  by fitting a least squares regression line for $\log\mathrm{MISE}$ versus $\log n$. The asymptotic rates of convergence under the four simulation scenarios are $O_p(n^{-1/2})$, $O_p(n^{-1/2})$, $O_p(n^{-2/5})$, and $O_p(n^{-2/5})$, respectively, as per \eqref{eq:rcge}, \eqref{eq:rcgp}, \eqref{eq:rcle}, and \eqref{eq:rclp} in Section \ref{sec:asym}. Since MISE in \eqref{eq:mise}  approximates the square distance between the true and fitted regression functions, theory predicts that the slopes of fitted least squares regression lines under the four simulation scenarios should be around $-1$, $-1$, $-0.8$, and $-0.8$, respectively, while the corresponding observed slopes were  $-0.99$, $-0.99$, $-0.8$, and $-0.8$. This remarkable agreement between theory and empirical behavior supports the relevance of the theory.


\subsection{Networks with Latent Block Structure}
\label{subsec:simuwsb}
To examine the performance of global and local network regression estimates on networks with latent block structure, we generate samples of networks from a weighted stochastic block model \citep{aicher2015learning}. Similar to Section \ref{subsec:simurates},  four different simulation scenarios I - IV are considered, corresponding to global and local network regression using both the Frobenius metric $d_F$ and the square root metric $d_{F, 1/2}$. Specifically, consider the weighted stochastic block model with the vector of community membership $z=(z_1, z_2)^\T$ where $z_i$ is a vector of length $m_i$ with all elements equal to $i$ and $m_1+m_2=m$. The existence of an edge between nodes of each block is governed by the Bernoulli distribution, parameterized by the corresponding entry in the matrix of edge probabilities $\theta=\begin{pmatrix}p_{11}&p_{12}\\p_{21}&p_{22}\end{pmatrix}$. The weights of existing edges are assumed to follow a beta distribution with shape parameters $\alpha = X$ and $\beta = 1-X$ for global network regression or $\alpha = \sin(\pi X)$ and $\beta = 1-\sin(\pi X)$ for local network regression. The associated graph Laplacian is then taken as the random response $L$ for the proposed regression models. For global network regression under the square root metric, the random response is taken as $F_2(L)$ to ensure that the linearity assumption is satisfied.

We investigated sample sizes $n=50, 100, 200, 500, 1000$, with $Q=1000$ Monte Carlo runs for each simulation scenario. In each iteration, random samples of pairs $(X_k, L_k)$, $k=1, \ldots, n$ were generated by sampling $X_k\sim\mathrm{U}(0, 1)$, setting $m_1=m_2=5$, $p_{11}=p_{22}=0.5$, $p_{12}=p_{21}=0.2$, and following the above procedure. The quality of the estimation for each simulation run was quantified by the integrated squared error as defined in Section \ref{subsec:simurates}. The bandwidths for the local network regression were chosen by leave-one-out cross-validation.

\begin{figure}[t]
\centering
\includegraphics[width=0.9\linewidth]{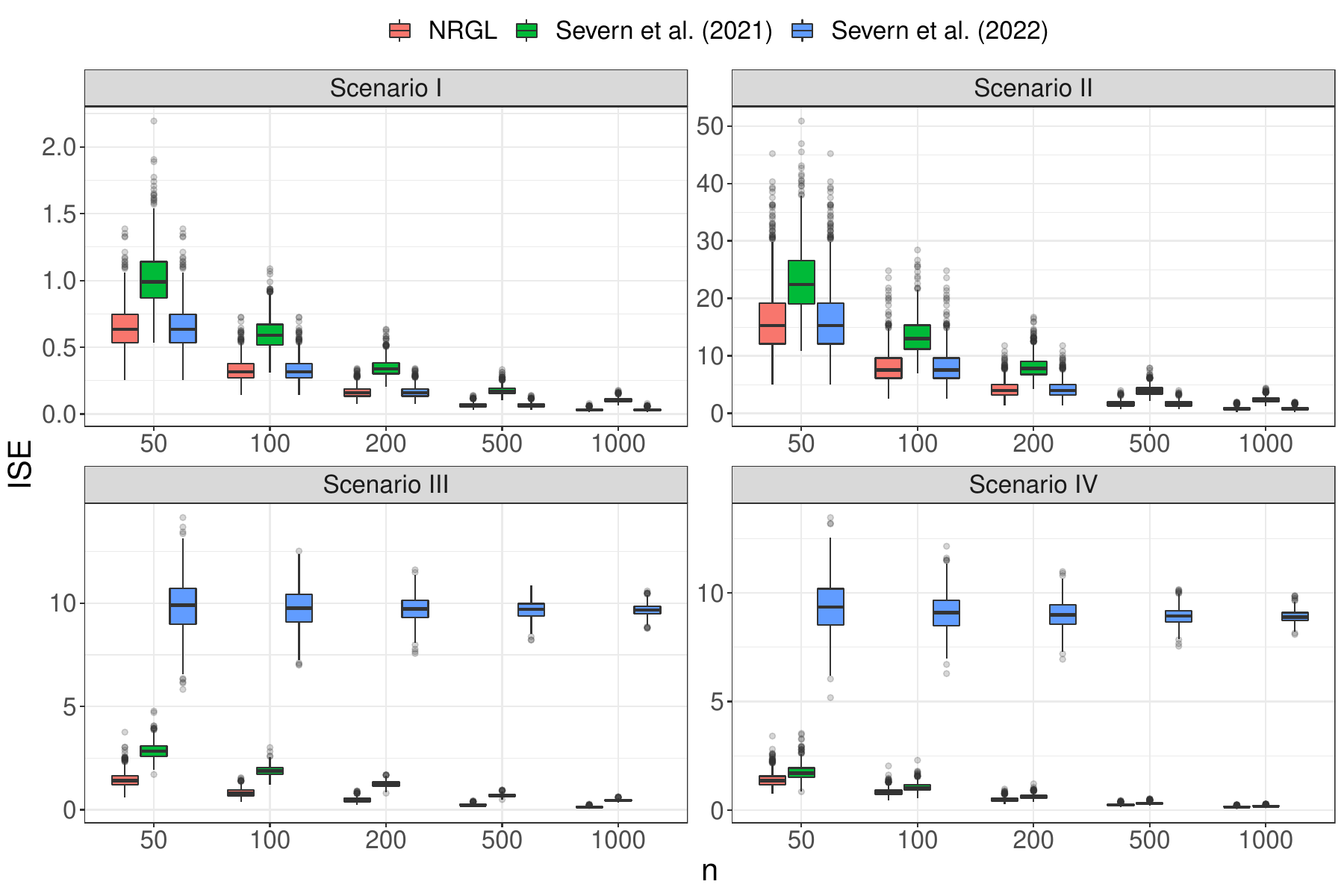}
\caption{Boxplots of integrated square errors (ISE) for networks with latent block structure using the proposed methods (NRGL, red) and the methods of \citet{severn2019manifold, severn2021non} (blue and green).}
\label{fig:simusbm}
\end{figure}

We also included comparisons with two alternative methods proposed by \citet{severn2019manifold} and \citet{severn2021non}. 
The integrated squared errors (ISE) for all simulation runs and different sample sizes under different simulation scenarios using the proposed methods and the two comparison methods are summarized in the boxplots in Figure~\ref{fig:simusbm}. The proposed network regression is  seen to perform as well as the method of \citet{severn2019manifold} under global scenarios and to achieve better performance compared to the methods in \citet{severn2019manifold, severn2021non} under local scenarios, especially for simulation scenario III.
Additional simulations for comparisons between different types of regression and metrics, and for networks generated from Erd\H{o}s-R\'{e}nyi random graph model \citep{erdHos1959random} are reported in Appendix C.

\section{Data Applications}
\label{sec:data}
\subsection{New York Yellow Taxi System After COVID-19 Outbreak}
\label{sec:taxi}
The yellow and green taxi trip records on pick-up and drop-off dates/times, pick-up and drop-off locations, trip distances, itemized fares, rate types, payment types and driver-reported passenger counts, collected by New York City Taxi and Limousine Commission (NYC TLC), are publicly available at \url{https://www1.nyc.gov/site/tlc/about/tlc-trip-record-data.page}. 
Additionally,  NYC Coronavirus Disease 2019 (COVID-19) data are available at \url{https://github.com/nychealth/coronavirus-data}, where one can find  citywide and borough-specific daily counts of probable and confirmed COVID-19 cases in New York City since February 29, 2020. This is the  date at which according to the Health Department the COVID-19 outbreak in NYC began. We aim to study the dependence of transport networks constructed from taxi trip records on covariates of interest, including COVID-19 new cases  and a weekend indicator, as travel patterns are well known to differ between weekdays and weekends. 

\begin{figure}[t]
\centering
\begin{subfigure}{.45\textwidth}
\centering
\includegraphics[width=\linewidth]{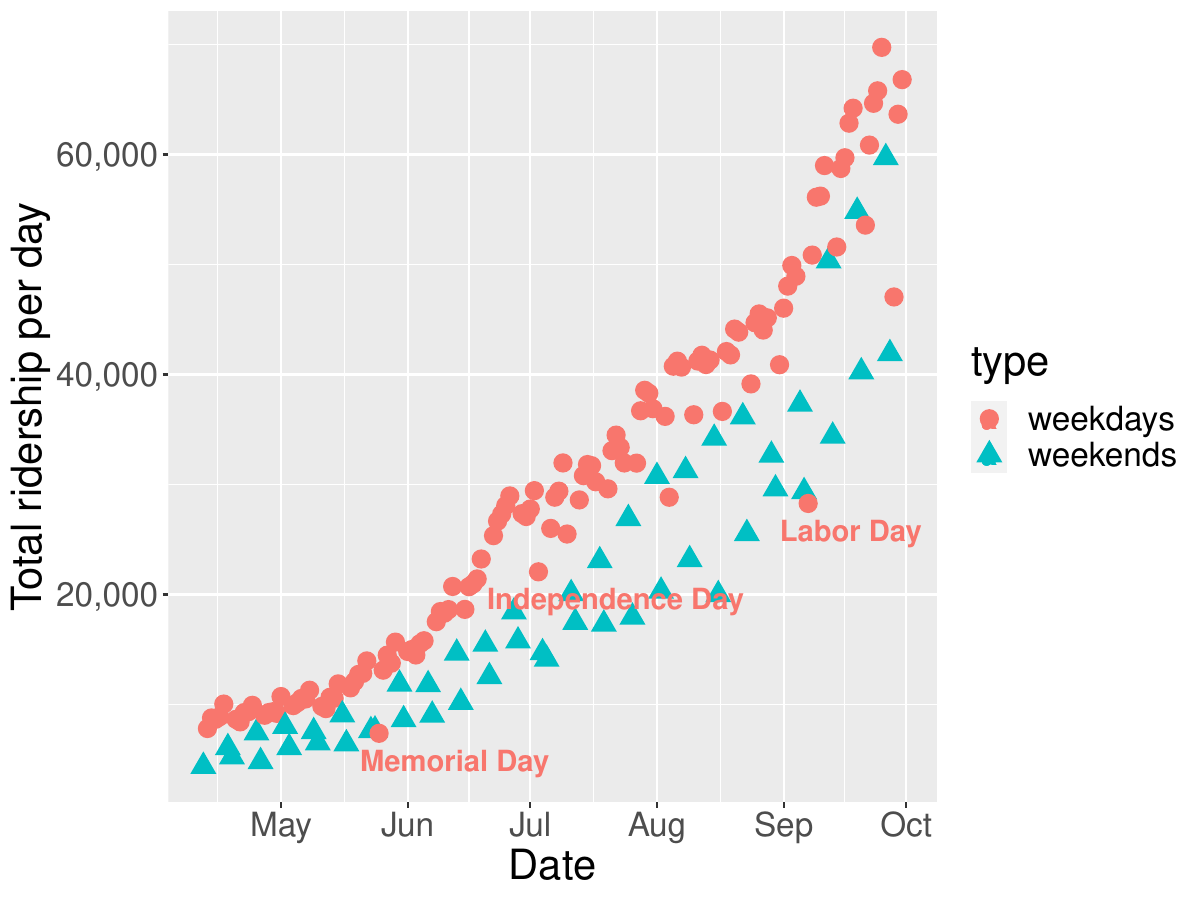}
\caption{}
\end{subfigure}%
\begin{subfigure}{.45\textwidth}
\centering
\includegraphics[width=\linewidth]{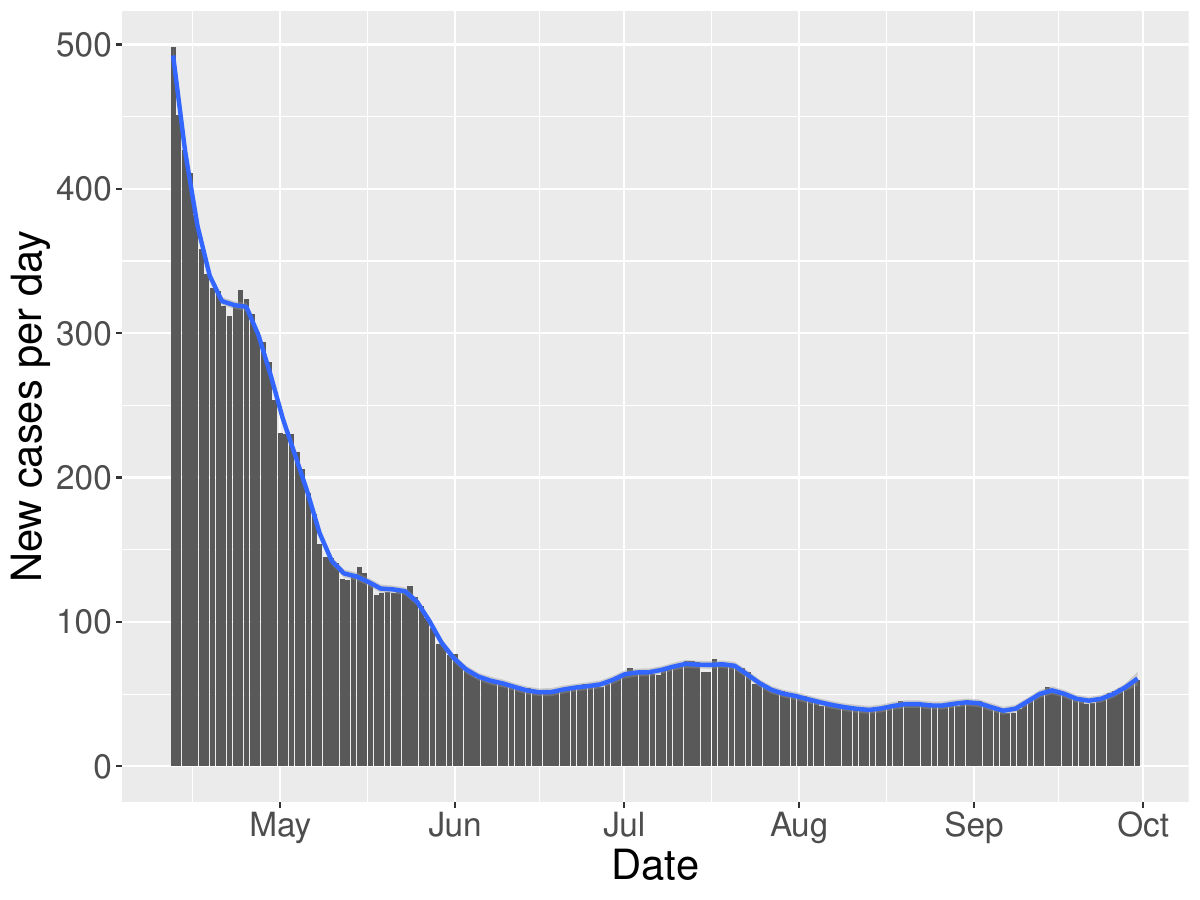}
\caption{}
\end{subfigure}
\begin{subfigure}{.45\textwidth}
\centering
\includegraphics[width=\linewidth]{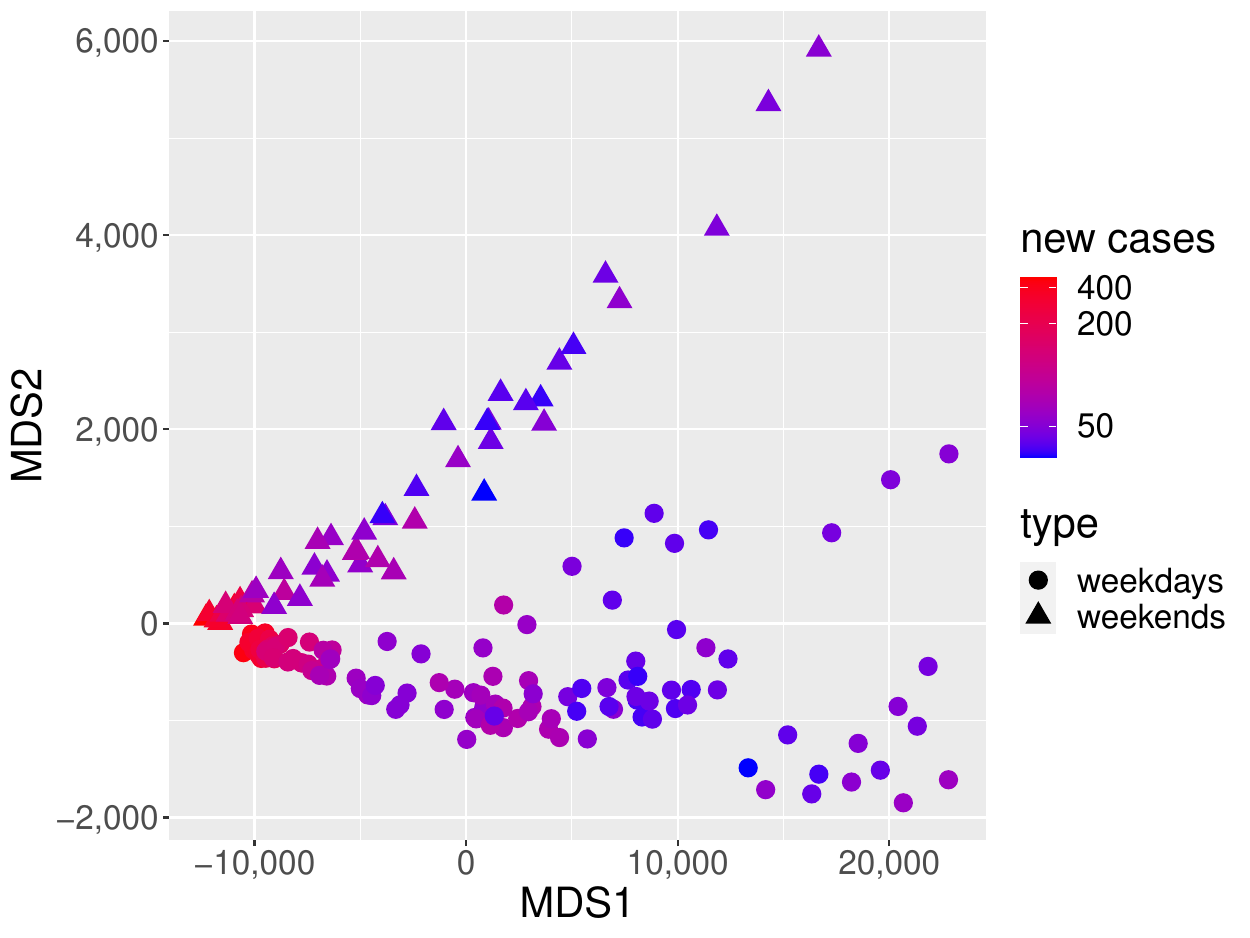}
\caption{}
\end{subfigure}%
\begin{subfigure}{.45\textwidth}
\centering
\includegraphics[width=\linewidth]{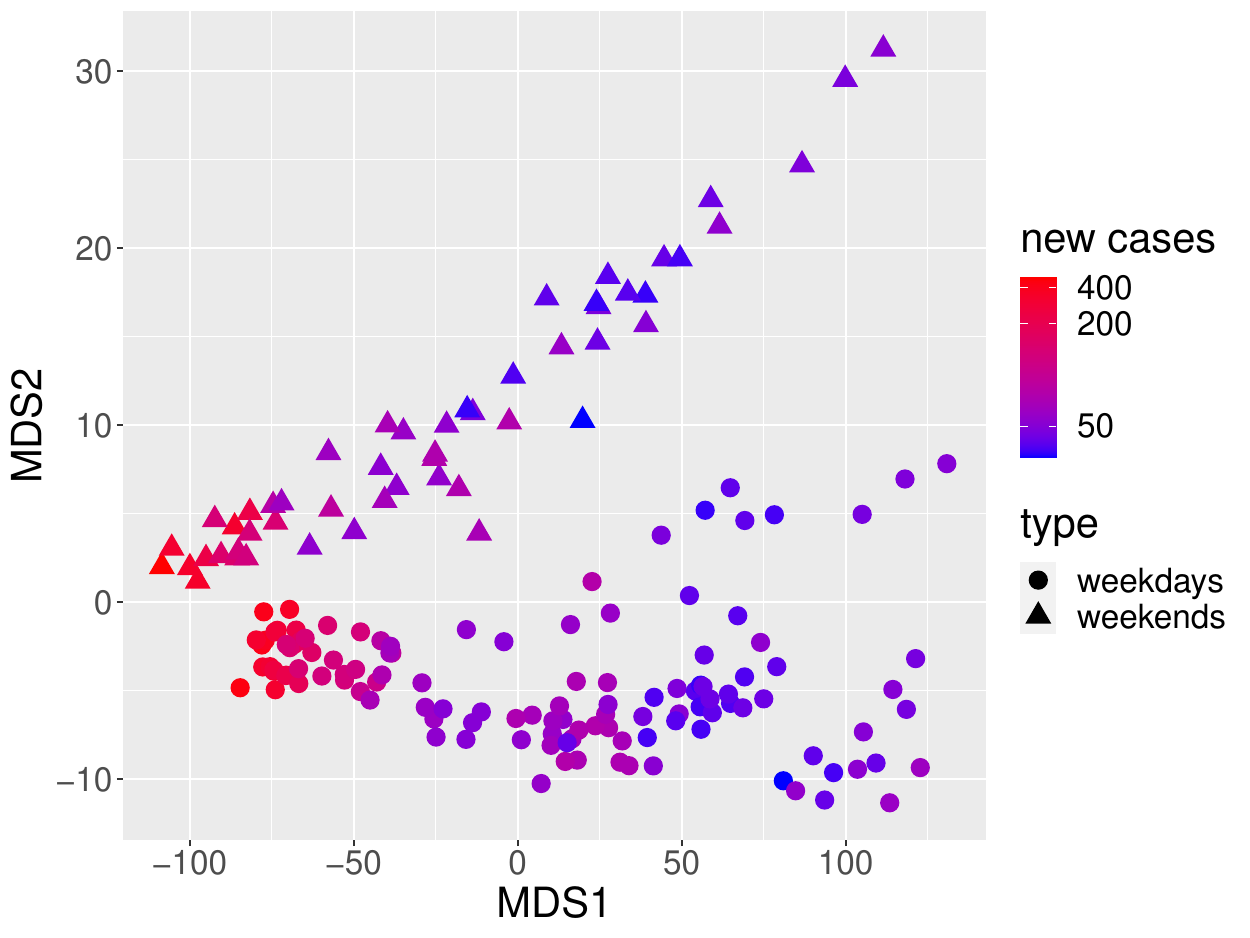}
\caption{}
\end{subfigure}
\caption{(a) Total yellow taxi ridership per day in Manhattan, New York in 2020, from April 12 to September 30.  Three holidays Memorial Day (May 25), Independence Day (July 3), and Labor Day (September 7) are highlighted. (b) COVID-19 new cases per day in Manhattan, New York in 2020, April 12 - September 30.   (c) MDS plot for taxi networks  based on the Frobenius metric $d_F$. (d) MDS plot for taxi networks based on the square root metric $d_{F, 1/2}$.}
\label{fig:mdsplot}
\end{figure}

We focused on yellow taxi trip records in the Manhattan area, which has the highest taxi traffic,  and grouped the $66$ taxi zones (excluding islands) as delimited by NYC TLC into $13$ regions. Details about the zones and regions are in Appendix D. Not long after the outbreak of COVID-19 in Manhattan, yellow taxi ridership, as measured by trips, began a steep decline during the week of March 15, reaching a trough around April 12. This motivated us to restrict our analysis to the time period comprising  the $172$ days from April 12, 2020 to September 30, 2020, during which yellow taxi ridership per day in Manhattan increased steadily. The total yellow taxi ridership per day is shown in Figure~\ref{fig:mdsplot}(a), where we observe a pronounced difference between weekends and weekdays. {Even though the three holidays Memorial Day (May 25), Independence Day (July 3), and Labor Day (September 7) are weekdays, they follow the same travel patterns as weekends, and consequently were classified as weekends in the following analyses.} 

For each day, we constructed a daily undirected network with nodes corresponding to the $13$ regions and edge weights representing the number of people who traveled between the regions connected by the edge on the specified day. Since the object of interest is the connection between different regions, we removed self-loops in the networks. We thus have observations that consist of  a simple undirected weighted network for each of the $172$ days from April 12 to September 30. Each of these networks is uniquely associated with a $13\times 13$ graph Laplacian. The covariates of interest include 
COVID-19 new cases for the day (see Figure~\ref{fig:mdsplot}(b))  and an indicator that is 1 for weekends and 0 otherwise.  
The first two multi-dimensional scaling (MDS) variables from a classical MDS analysis of the resulting  graph Laplacians provide exploratory analysis and are  shown in Figure~\ref{fig:mdsplot}(c) and \ref{fig:mdsplot}(d). These MDS plots indicate  that irrespective of the chosen  metric, there is a clear separation between weekdays and weekends in MDS2 and also that the number of COVID-19 new cases plays an important role in MDS1.

Since the weekend indicator is a binary predictor, we applied global network regression, using both the Frobenius metric $d_F$ and the square root metric $d_{F, 1/2}$. The estimated mean squared prediction error (MSPE) was calculated for each metric using ten-fold cross validation, averaged over $100$ runs. A common baseline  metric, for which we chose the Frobenius metric $d_F$, was used to calculate the MSPEs to make them comparable across metrics. The MSPE for $d_{F, 1/2}$ was found to be $96.4\%$ of that for $d_{F}$, validating the utility of the square root metric in real-world applications. Additionally, the \f coefficient of determination $R_{\oplus}^2=1-E\{d^2[L, m_{G}(x)]\}/V_{\oplus}$, an extension of the coefficient of determination $R^2$ for linear regression, can be similarly used to quantify the proportion of response variation ``explained'' by the covariates. The corresponding sample version is $\hat{R}_{\oplus}^2=1-\sum d^2[L_k, \hat{m}_{G}(X_k)]/\sum d^2(L_k, \hat{\omega}_{\oplus})$, where $\hat{\omega}_{\oplus}=\argmin_{\omega\in\mathcal{L}_m}\sum d^2(L_k, \omega)$. We found that  $\hat{R}_{\oplus}^2=0.433$ for $d_F$ and $\hat{R}_{\oplus}^2=0.453$ for $d_{F, 1/2}$, which further lends support for the use of $d_{F, 1/2}$ in this specific application. Also, we observe that the generalized tensor-response regression of \citet{zhang2018generalized} can be applied to these data  using  a log link, as the edge weights  are counts. We compared this approach with the proposed  global network regression for the square root metric $d_{F, 1/2}$. The MSPE of the  proposed network regression averaged over $100$ runs was substantially smaller by a factor 0.51 of that for tensor-response regression, clearly favoring the proposed method.     

\begin{figure}[t]
\centering
\begin{subfigure}{.45\textwidth}
\centering
\includegraphics[width=\linewidth]{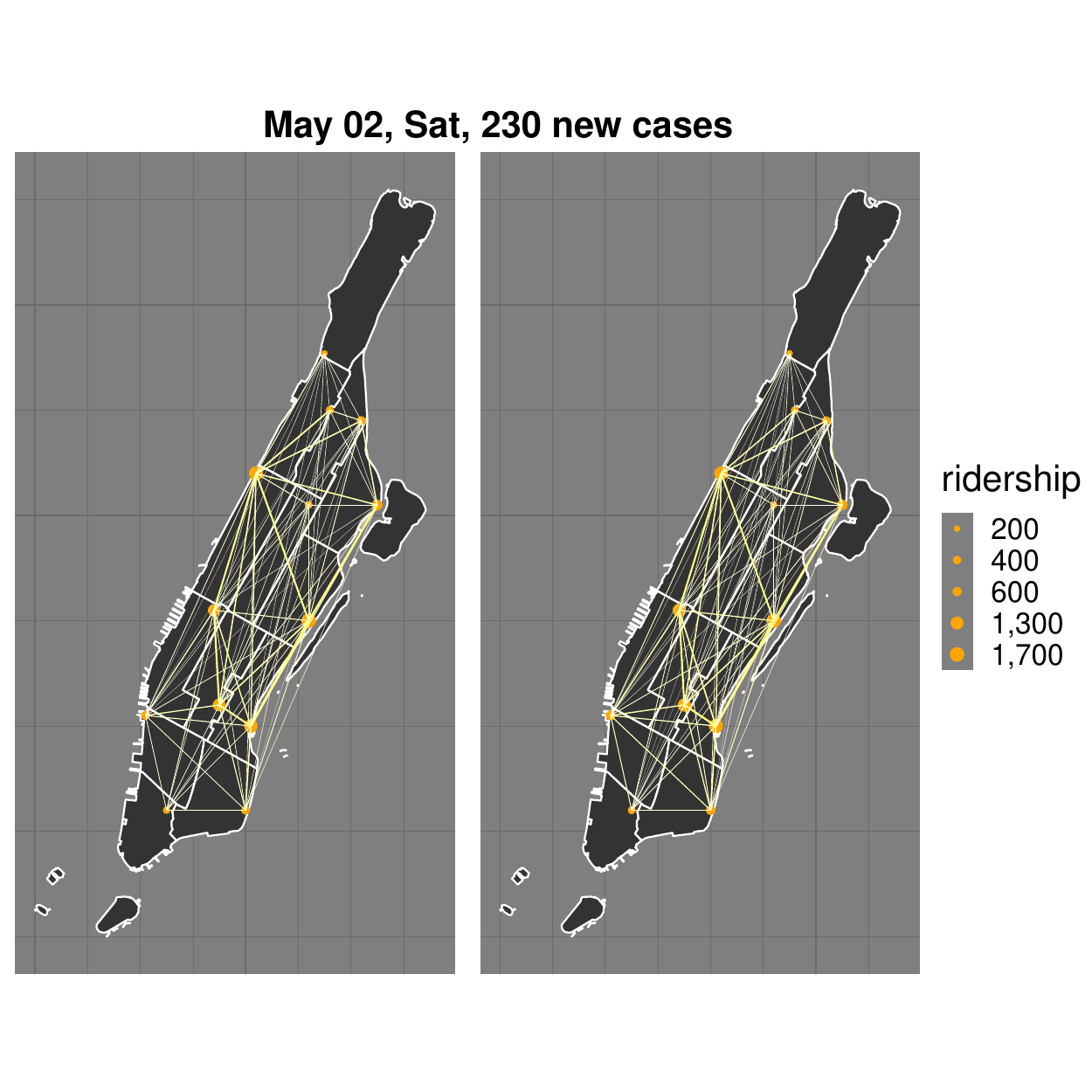}
\end{subfigure}%
\begin{subfigure}{.45\textwidth}
\centering
\includegraphics[width=\linewidth]{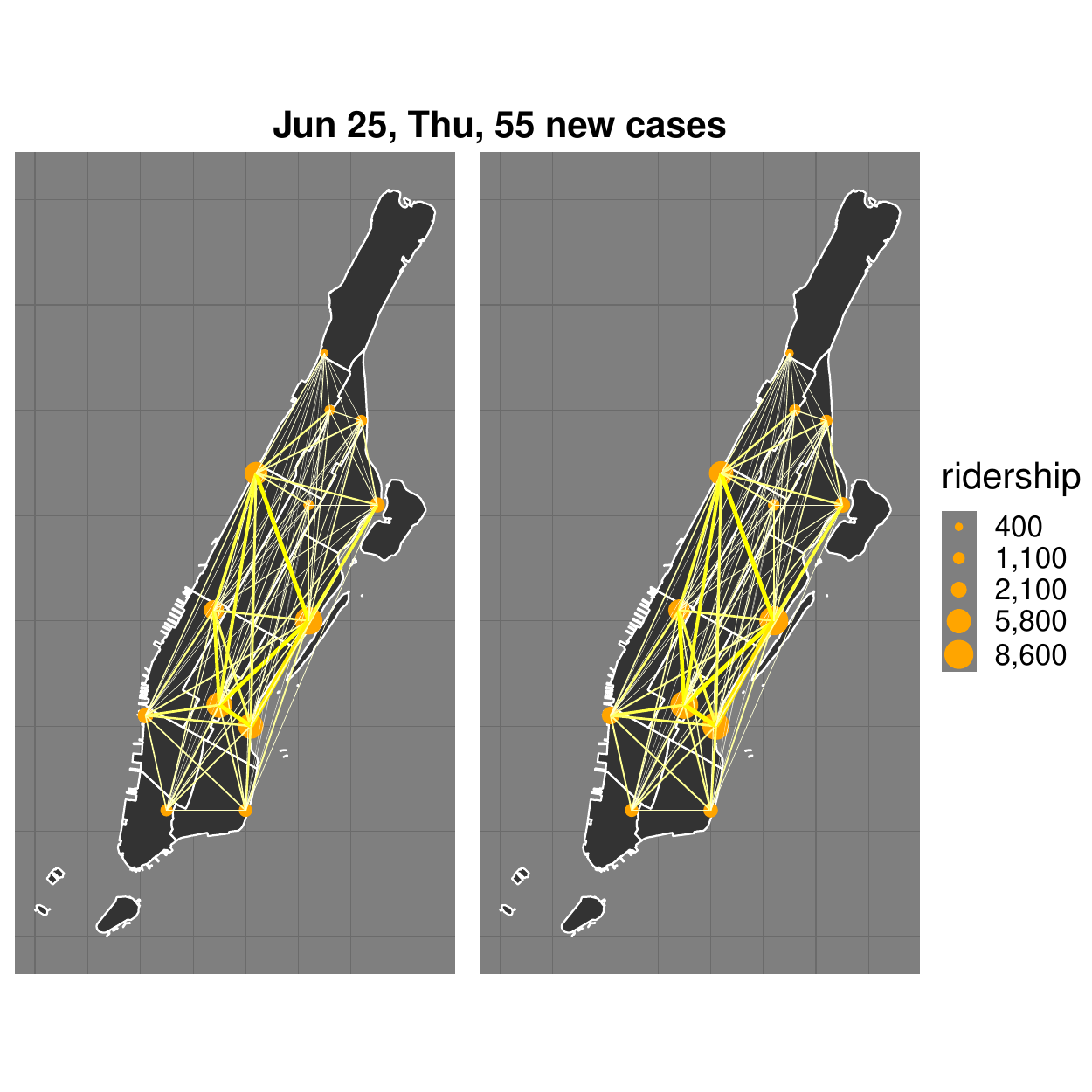}
\end{subfigure}
\begin{subfigure}{.45\textwidth}
\centering
\includegraphics[width=\linewidth]{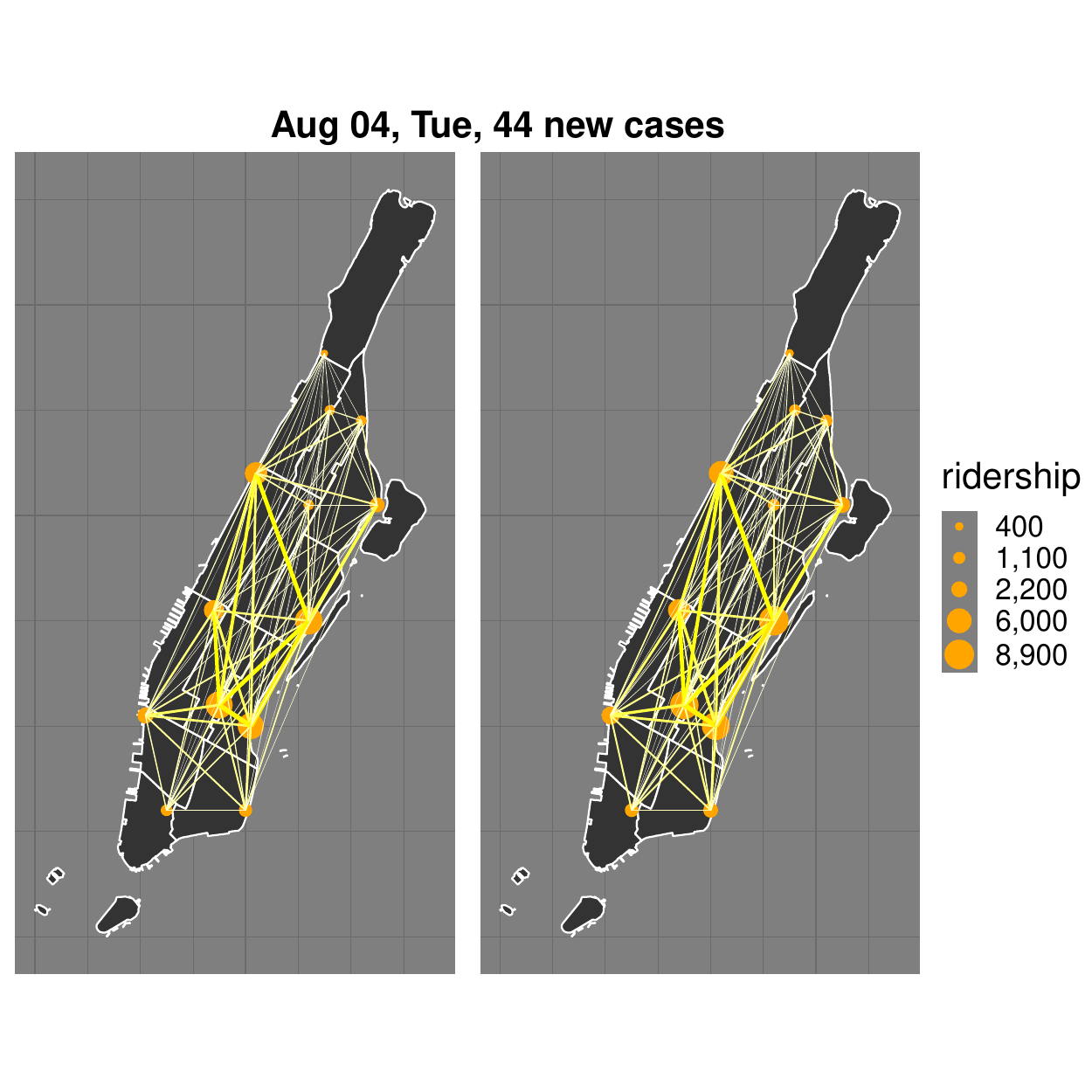}
\end{subfigure}%
\begin{subfigure}{.45\textwidth}
\centering
\includegraphics[width=\linewidth]{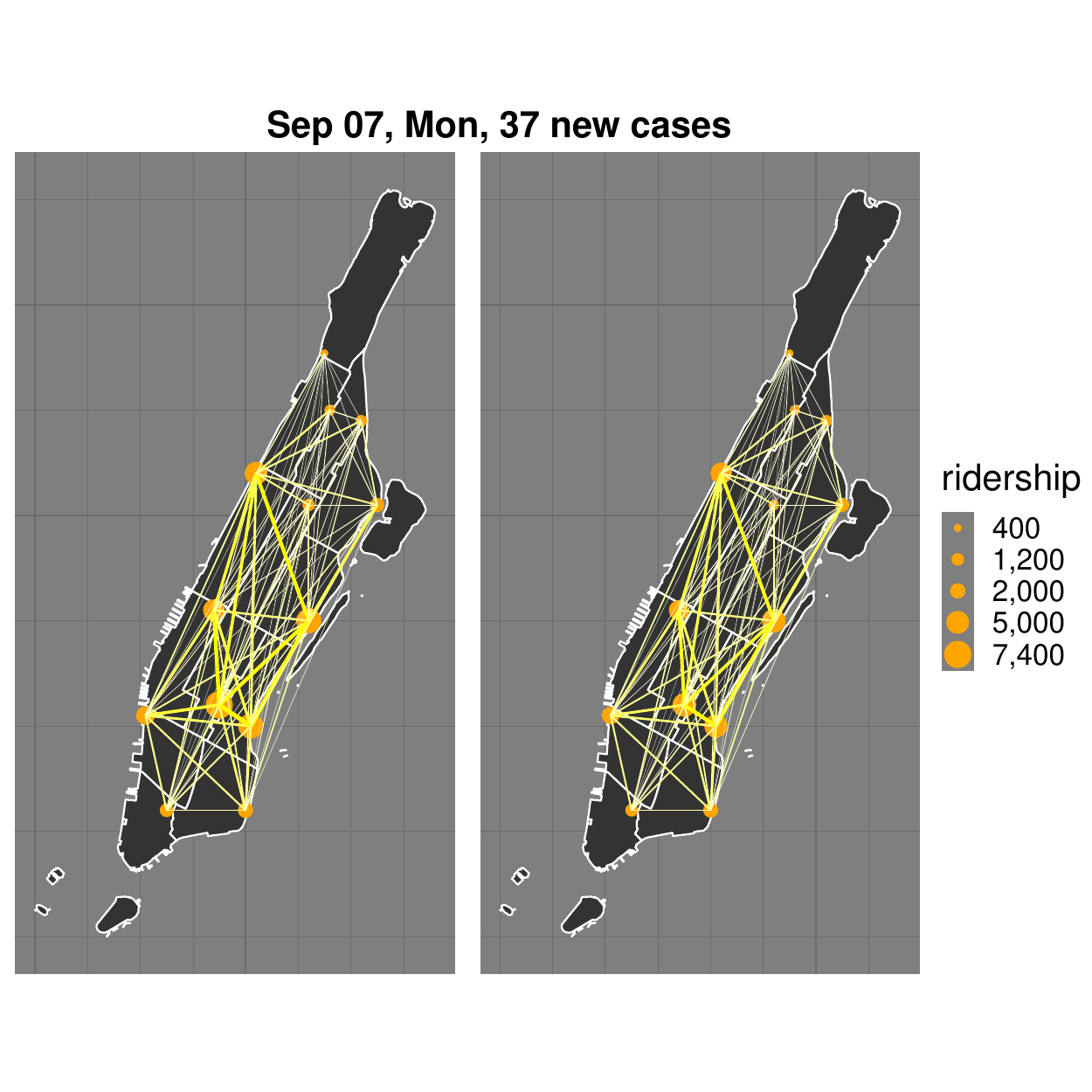}
\end{subfigure}
\caption{True (left) and fitted (right) networks on May 2, Jun 25, Aug 4,  and Sep 7, 2020  (from top left to bottom right). The corresponding days  and the number of COVID-19 new cases are in the headline of each subfigure.}
\label{fig:taxifitted}
\end{figure}

True and fitted networks using the square root metric $d_{F, 1/2}$ for four selected days  are shown in Figure~\ref{fig:taxifitted}. From top left to bottom right, the four days  were chosen to be spaced evenly in the considered time period April 12 to September 30. 
For each day, on the left is the observed and on the right the fitted network as obtained from network regression.  The size of the nodes indicates the volume of traffic in this region and the thickness of the edges represents their weights. The fitted network regression  model is seen to capture both structure and weight information of the networks given relevant covariates, and  demonstrates trends that are in line with observations.

\begin{figure}[tbp]
	\centering
	\includegraphics[width=0.84\linewidth]{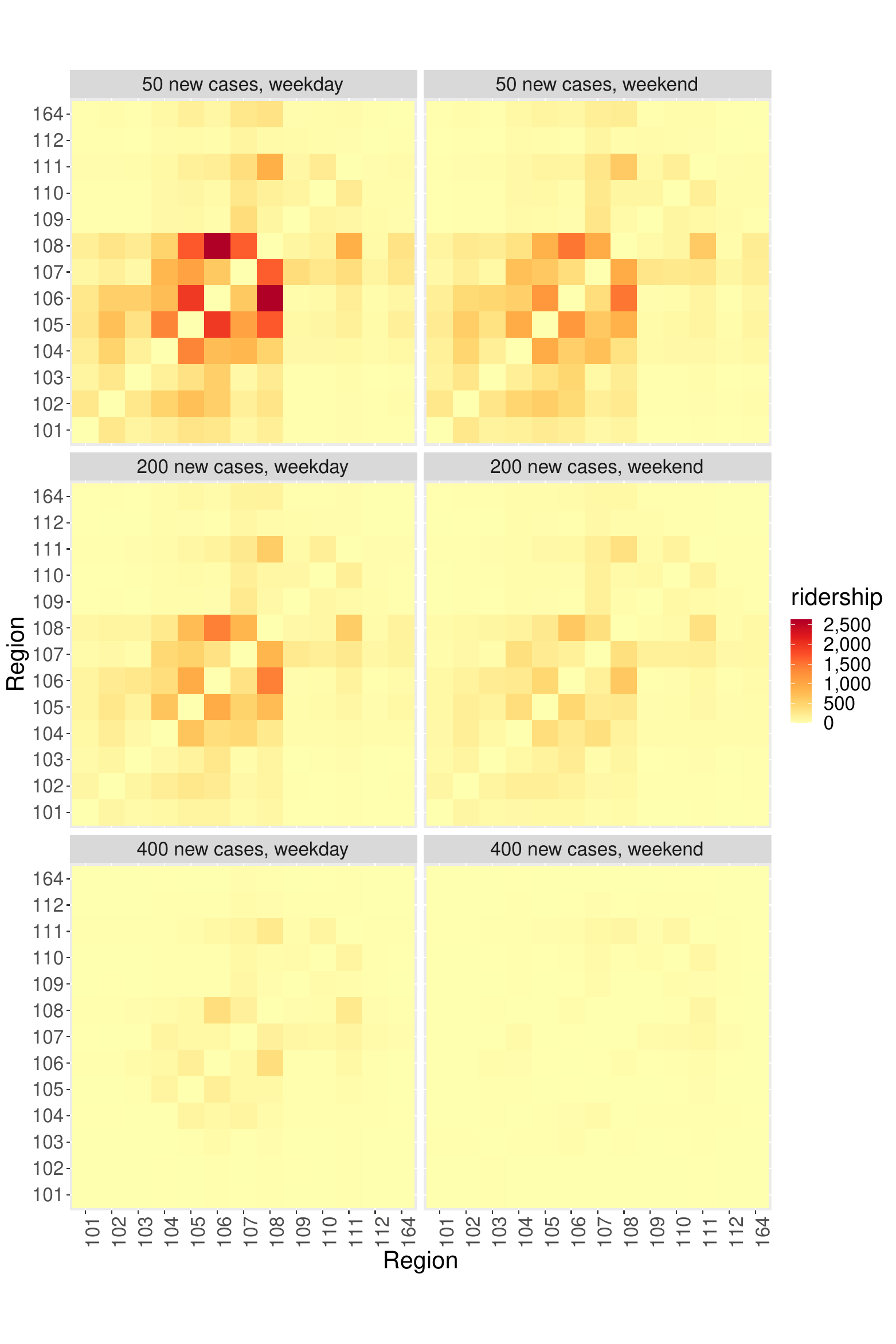} \vspace{-.5cm}
	\caption{Predicted networks represented as heatmaps at different number of COVID-19 new cases on weekdays or weekends. The top, middle, and bottom rows show, respectively, the predicted networks at $50$, $200$, and $400$ COVID-19 new cases. The left and right columns depict the predicted networks on weekdays and weekends, respectively.}
	\label{fig:taxicovidinterpretfacet}
\end{figure}

To further investigate the effects of COVID-19 new cases and weekends,  predicted networks represented as heatmaps at $50$, $200$, and $400$ COVID-19 new cases for weekdays or weekends are shown in Figure~\ref{fig:taxicovidinterpretfacet}. Edge weights are seen to decrease for increasing COVID-19 new cases, reflecting the negative impact of the epidemic on travel. Heatmaps are increasingly concentrated  with increasing numbers of new cases of COVID-19, indicating a narrowing of   movements to limited areas. 
Weekend taxi traffic, with lighter and less essential traffic compared to weekdays, is more severely affected by COVID-19 and as new cases approach 400 per day comes to a near stop.  The regions with the heaviest traffic are areas  105, 106, 107, and 108, which are chiefly residential areas and include popular locations such as Penn Station, Grand Central Terminal, and also the Metropolitan Museum of Art. 

\begin{figure}[tbp]
\centering
\includegraphics[width=0.80\linewidth]{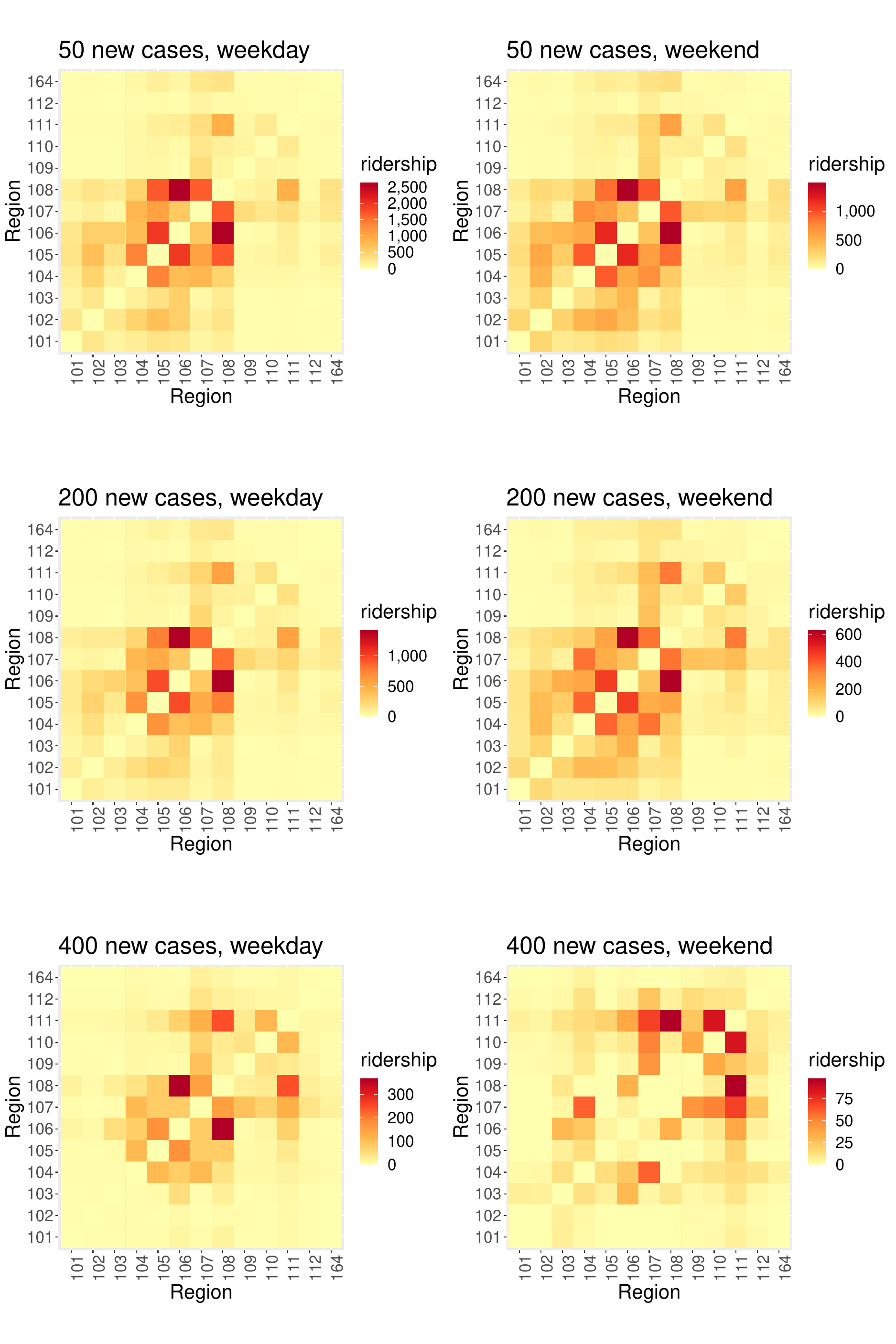}
\caption{Predicted networks represented as heatmaps at different number of COVID-19 new cases on weekdays or weekends. Each heatmap has its own scale to enhance visualization of structural changes in connections in dependence on daily COVID-19 new cases.  The top, middle, and bottom rows show, respectively, the predicted networks at $50$, $200$, and $400$ COVID-19 new cases. The left and right columns depict the predicted networks on weekdays and weekends, respectively.}
\label{fig:taxicovidinterpret}
\end{figure}

To further illustrate   the effects of COVID-19 new cases and weekends versus weekdays on network structure, Figure~\ref{fig:taxicovidinterpret} shows the same predicted networks where now each heatmap has its own  color scale. This indicates  that higher case numbers lead to bigger structural changes in traffic patterns  on  weekends  than on weekdays, likely because weekend travel tends to be optional. 
Blocks involving regions 101, 102, 103 have declining traffic with increasing COVID-19 new cases on both weekdays and weekends; 
these regions are in lower Manhattan, the central borough for business (see Appendix D), which includes the Financial District and the World Trade Center. This likely reflects that more people work from home with increasing case numbers and demonstrates the flexibility of the fits obtained from the proposed network regression. 

\subsection{Dynamics of Networks in the Aging Brain}
\label{sec:neuro}
The increasing availability of neuroimaging data, such as functional magnetic resonance imaging (fMRI) data, has facilitated the investigation of age-related changes in human brain network organization. Resting-state fMRI (rs-fMRI), as an important modality of fMRI data acquisition, has been widely used to study normal aging, which is known to be associated with cognitive decline, even in individuals without any process of retrogressive disorder \citep{ferreira2013resting, sala2014changes, sala2015reorganization}.

FMRI measures brain activity by detecting changes in blood-oxygen-level-dependent (BOLD) signals in the brain across time. During recordings of rs-fMRI 
subjects relax during the sequential acquisition of fMRI scans. Spontaneous fluctuations in brain activity during rest is reflected by low-frequency oscillations of the BOLD signal, recorded as voxel-specific time series of activation strength. Network-based analyses of brain functional connectivity at the subject level typically rely on a specific spatial parcellation of  the brain into a set of regions of interest (ROIs) \citep{bullmore2009complex}. Temporal coherence between pairwise ROIs is usually measured by so-called temporal Pearson correlation coefficients (PCC) of the fMRI time series, forming a $m\times m$ correlation matrix when considering $m$ distinct ROIs. This correlation matrix assumes the role of observed functional connectivity for each subject. 
Hard or soft thresholding \citep{schwarz2011negative} is customarily applied to produce a binary or weighted functional connectivity network.

Data used in our study were obtained from the Alzheimer’s Disease Neuroimaging Initiative (ADNI) database (\url{adni.loni.usc.edu}), where $n=404$ cognitively normal elderly subjects with age ranging from 55.61 to 95.39 years participated in the study; one rs-fMRI scan is randomly selected if multiple scans are available for a subject. We used the automated anatomical labeling (AAL) atlas \citep{tzourio2002automated} to parcellate the whole brain into $90$ ROIs, with $45$ ROIs in each hemisphere. Details about the ROIs can be found in Appendix D. Preprocessing was carried out in MATLAB using the Statistical Parametric Mapping (SPM12, \url{www.fil.ion.ucl.ac.uk/spm}) and Resting-State fMRI Data Analysis Toolkit V1.8 (REST1.8, \url{http://restfmri.net/forum/?q=rest}). Briefly, this included the removal of any artifacts from head movement, correction for differences in image acquisition time between slices, normalization to the standard space, spatial smoothing and temporal filtering (bandpass filtering of 0.01-0.1 Hz). The mean time course of the voxels within each ROI was then extracted for network construction. A PCC matrix was calculated for all time course pairs for each subject. These matrices were then converted into simple, undirected, weighted networks by setting diagonal entries to $0$ and thresholding the absolute values of the remaining correlations. 
We used density-based thresholding \citep{fornito2016fundamentals}, where the threshold is allowed to vary from subject to subject to achieve a desired, fixed connection density. Specifically, in our analyses the $15\%$ strongest connections were kept.

\begin{figure}[t]
\centering
\includegraphics[width=0.9\linewidth]{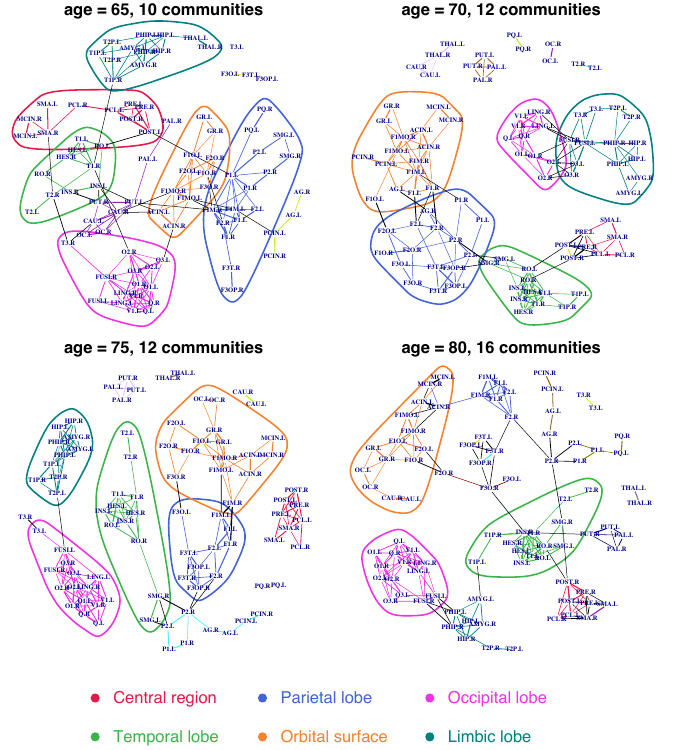}
\caption{Topological representation using spectral community detection for predicted functional connectivity networks at different ages (in years). The communities comprising 10 or more ROIs are highlighted using colored polygons. These communities are found to be associated with different anatomical regions of the brain (see Table~\ref{tab:aal} in Appendix D).}
\label{fig:adnicommunitylabel}
\end{figure}

To investigate age-related changes in human brain network organization, we employed local network regression using the Frobenius metric $d_F$ with graph Laplacians corresponding to the networks constructed from PCC matrices as responses, with age (in years) as scalar-valued covariate. The bandwidth for the predictor age was chosen to minimize the prediction error using leave-one-out cross-validation, resulting in a bandwidth of $h=0.20$. Prediction was performed at four different ages: $65, 70, 75, $ and $80$ (approximately the $20\%, 40\%, 60\%$, and $80\%$ quantiles of the age distribution of the $404$ subjects). The predicted networks are demonstrated in Figure~\ref{fig:adnicommunitylabel}, where the nodes were placed using the Fruchterman-Reingold layout algorithm \citep{fruchterman1991graph} for visualization. Spectral clustering \citep{newman2006finding} was applied to detect the community structure in each network, where different communities are distinguished by different colors. The number of communities for ages $65$, $70$, $75$, and $80$ was $10$, $12$, $12$, and $16$, respectively. The communities with no less than $10$ nodes are highlighted using colored polygons. These communities are found to be associated with different anatomical regions of the brain (see Table~\ref{tab:aal} in Appendix D), where a community is identified as the anatomical region to which the majority of nodes belong. 
As can be seen in Figure~\ref{fig:adnicommunitylabel}, the communities associated with the central region, the parietal lobe, and the limbic lobe disintegrate into several small communities with increasing age. This finding suggests that higher age is associated with  increased local interconnectivity and cliquishness. High cliquishness is known to be associated with reduced capability to rapidly combine specialized information from distributed brain regions, which may contribute to cognitive decline for healthy elderly adults \citep{sala2015reorganization}.

\section{Discussion}
\label{sec:dis}
The proposed network regression models provide a  novel technology for analyzing network objects, with extensive applications in various areas including neuroimaging and social sciences. We present theoretical justifications that include rates of convergence for both global and local versions. The pointwise rates of convergence are optimal for both global and local versions in the sense that they correspond to the known optimal rates in the special case of Euclidean objects. In the proposed framework, the number of nodes $m$ is assumed to be fixed.  If $m\to\infty$, the theoretical results will no longer hold and the asymptotics for this case would be a topic for future research. 

Our framework can be easily extended to the case of directed networks. For a directed network $G$, as long as $G$ is simple, i.e., there are no self-loops or multi-edges, one has a  one-to-one correspondence between $G$ and its graph Laplacian. Therefore we can still represent networks using their corresponding graph Laplacians. The only difference is that the graph Laplacian is no longer symmetric. As such, the space of graph Laplacians $\mathcal{L}_m$ is then of dimension $m(m-1)$, rather than $m(m-1)/2$. However, $\mathcal{L}_m$ is still convex and closed, which ensures the existence and uniqueness of projections onto $\mathcal{L}_m$, and  asymptotic properties can be derived by arguments that closely follow those provided in this paper. 

For the case of a  time series of networks, if one adopts an autoregressive model both the response and the predictor are network objects; see for example \citet{jian:20}. This case is beyond the scope of the present paper.  One potential approach is to use geodesics in the space of graph Laplacians. The relationship between response and predictor networks can then be studied by relating movements along geodesics \citep{zhu2021autoregressive}.

\section*{Acknowledgments}
\label{sec:akd}
We thank Michael Mahoney and three anonymous reviewers for their helpful comments which have helped to improve this article. Data used in preparation of this article were obtained from the Alzheimer's Disease Neuroimaging Initiative (ADNI) database (\url{adni.loni.usc.edu}). As such, the investigators within the ADNI contributed to the design and implementation of ADNI and/or provided data but did not participate in analysis or writing of this report. A complete listing of ADNI investigators can be found at: \url{http://adni.loni.usc.edu/wp-content/uploads/how_to_apply/ADNI_Acknowledgement_List.pdf}. Data collection and sharing for this project was funded by the Alzheimer’s Disease Neuroimaging Initiative (ADNI) (National Institutes of Health Grant U01 AG024904) and DOD ADNI (Department of Defense award number W81XWH-12-2-0012). This research was supported in part by NSF grant DMS-2014626.


\appendix
\section*{Appendix A. Assumptions for Theorem \ref{thm:euclidean:l} and Theorem \ref{thm:euclideanpower:l}}
In the following, $f_X(\cdot)$ and $f_{X|L}(\cdot, \omega)$ stand for the marginal density of $X$ and the conditional density of $X$ given $L=\omega$, respectively. $\mathcal{T}$ is a closed interval in $\mathbb{R}$ with interior $\mathcal{T}^o$.
\begin{enumerate}[label=(A\arabic*), leftmargin=1cm]
	\item The kernel $K(\cdot)$ is a probability density function, symmetric around zero. Furthermore, defining $K_{kj}=\int_{\mathbb{R}}K^k(u)u^jdu$, $|K_{14}|$ and $|K_{26}|$ are both finite.\label{itm:lp0}
	\item $f_X(\cdot)$ and $f_{X\mid L}(\cdot, \omega)$ both exist and are twice continuously differentiable, the latter for all $\omega\in\mathcal{L}_m$, and $\sup_{x, \omega}|(\partial^2 f_{X\mid L}/\partial x^2)(x, \omega)|<\infty$. Additionally, for any open set $U\subset\mathcal{L}_m$, $\int_UdF_{L\mid X}(x, \omega)$ is continuous as a function of $x$.\label{itm:lp1}
	\item The kernel $K(\cdot)$ is a probability density function, symmetric around zero, and uniformly continuous on $\mathbb{R}$. Furthermore, defining $K_{jk}=\int_{\mathbb{R}}K(u)^ju^kdu$ for $j, k\in\mathbb{N}$, $|K_{14}|$ and $|K_{26}|$ are both finite. The derivative $K'$ exists and is bounded on the support of $K$, i.e., $\sup_{K(x)>0}|K'(x)|<\infty$; additionally, $\int_{\mathbb{R}}x^2|K'(x)|(|x\log|x||)^{1/2}dx<\infty$.\label{itm:lu0}
	\item $f_X(\cdot)$ and $f_{X\mid L}(\cdot, \omega)$ both exist and are continuous on $\mathcal{T}$ and twice continuously differentiable on $\mathcal{T}^o$, the latter for all $\omega\in\mathcal{L}_m$. The marginal density $f_X(\cdot)$ is bounded away from zero on $\mathcal{T}$, $\inf_{x\in\mathcal{T}}f_X(x)>0$. The second-order derivative $f''_X$ is bounded, $\sup_{x\in\mathcal{T}^o}|f''_X(x)|<\infty$. The second-order partial derivatives $(\partial^2 f_{X\mid L}/\partial x^2)(\cdot, \omega)$ are uniformly bounded, $\sup_{x\in\mathcal{T}^o, \omega\in\mathcal{L}_m}|(\partial^2 f_{X\mid L}/\partial x^2)(x, \omega)|<\infty$. Additionally, for any open set $U\subset\mathcal{L}_m$, $\int_UdF_{L\mid X}(x, \omega)$ is continuous as a function of $x$;  $M(\cdot, x)$ is equicontinuous, i.e., for all  $x\in\mathcal{T}$,
	\[\limsup_{z\to x}\sup_{\omega\in\mathcal{L}_m}|M(\omega, z)-M(\omega, x)|=0.\]\label{itm:lu1}
\end{enumerate}

\section*{Appendix B. Proofs}
\label{supp:proof}
\subsection*{B.1 Conditions}
\label{supp:condition}
To obtain rates of convergence for the global and local network regression estimators, we require the following conditions that parallel those in  \citet{mull:19:3} and \citet{chen2020uniform}. For ease of presentation, we replace the graph Laplacian $L$ in global and local network regression by a general random object $Y$ taking values in an arbitrary metric space $(\Omega, d)$ and follow the same notations there. Consequently, the following conditions apply to any random objects taking values in a metric space. We will verify that the two metric spaces, $(\mathcal{L}_m, d_F)$ and $(\mathcal{M}_m, d_F)$, satisfy these conditions, so that we indeed can apply these general conditions and ensuing results. This lays the foundation for the derivation of rates of convergence for the corresponding estimators.

The following conditions are required to obtain consistency and rates of convergence of $\hat{m}_{G}(x)$. For a fixed $x\in\mathbb{R}^p$:
\begin{enumerate}[label=(B\arabic*), leftmargin=1cm]
	\item 
	The objects $m_{G}(x)$ and $\hat{m}_{G}(x)$ exist and are unique, the latter almost surely, and, for any $\varepsilon>0$, 
	\[\inf_{d[m_{G}(x), \omega]>\varepsilon}M_G(\omega, x)>M_G[m_{G}(x), x].\]\label{itm:gp0}
	\item 
	Let $B_\delta[m_{G}(x)]\subset\Omega$ be the ball of radius $\delta$ centered at $m_{G}(x)$ and $N\{\varepsilon, B_\delta[m_{G}(x)], d\}$ be its covering number using balls of size $\varepsilon$. Then
	\[\int_0^1(1+\log N\{\delta\varepsilon, B_{\delta}[m_{G}(x)], d\})^{1/2}d\varepsilon=O(1)\quad\mbox{ as }\delta\to0.\]\label{itm:gp1}
	\item
	There exists $\eta_0>0$, $C_0>0$ and $\gamma_0>1$, possibly depending on $x$, such that
	\[\inf_{d[m_{G}(x), \omega]<\eta_0}\{M_G(\omega, x)-M_{G}[m_{G}(x), x]-C_0d[m_{G}(x), \omega]^{\gamma_0}\}\geq 0.\]\label{itm:gp2}
\end{enumerate}
Uniform convergence results require stronger versions of the above conditions. Let $\|\cdot\|_E$ be the Euclidean norm on $\mathbb{R}^p$ and $B>0$ a given constant.
\begin{enumerate}[label=(B\arabic*), leftmargin=1cm]
	\setcounter{enumi}{3}
	\item
	Almost surely, for all $\|x\|_E<B$, the objects $m_{G}(x)$ and $\hat{m}_{G}(x)$ exist and are unique. Additionally, for any $\varepsilon>0$, 
	\[\inf_{\|x\|_E\leq B}\inf_{d[m_{G}(x), \omega]>\varepsilon}\{M_G(\omega, x)-M_G[m_{G}(x), x]\}>0\] and there exists $\zeta=\zeta(\varepsilon)>0$ such that \[\p(\inf_{\|x\|_E\leq B}\inf_{d[\hat{m}_{G}(x), \omega]>\varepsilon}\{\hat{M}_{G}(\omega, x)-\hat{M}_{G}[\hat{m}_{G}(x), x]\}\geq\zeta)\to1.\]\label{itm:gu0}
	\item
	With $B_\delta[m_{G}(x)]$ and $N\{\varepsilon, B_\delta[m_{G}(x)], d\}$ as in Condition \ref{itm:gp1},
	\[\int_0^1\sup_{\|x\|_E\leq B}(1+\log N\{\delta\varepsilon, B_{\delta}[m_{G}(x)], d\})^{1/2}d\varepsilon=O(1)\quad\mbox{ as }\delta\to0.\]\label{itm:gu1}
	\item
	There exist $\tau_0>0$, $D_0>0$, and $\rho_0>1$, possibly depending on $B$, such that
	\[\inf_{\|x\|_E\leq B}\inf_{d[m_{G}(x), \omega]<\tau_0}\{M_G(\omega, x)-M_G[m_{G}(x), x]-D_0d[m_{G}(x), \omega]^{\rho_0}\}\geq0.\]\label{itm:gu2}
\end{enumerate}
We require the following conditions to obtain pointwise rates of convergence for $\hat{m}_{L, n}(x)$. For simplicity, we assume that the marginal density $f_X(\cdot)$ of $X$ has unbounded support, and consider $x\in\mathbb{R}$ with $f_X(x)>0$.
\begin{enumerate}[label=(B\arabic*), leftmargin=1cm]
	\setcounter{enumi}{6}
	\item
	The minimizers $m(x), m_{L, h}(x)$ and $\hat{m}_{L, n}(x)$ exist and are unique, the last almost surely. Additionally, for any $\varepsilon>0$,
	\[\inf_{d[m(x), \omega]>\varepsilon}\{M(\omega, x)-M[m(x), x]\}>0,\]
	\[\liminf_{h\to0}\inf_{d[m_{L, h}(x), \omega]>\varepsilon}\{M_{L, h}(\omega, x)-M_{L, h}[m_{L, h}(x), x]\}>0.\]\label{itm:lp2}
	\item
	Let $B_\delta[m(x)]\subset\Omega$ be the ball of radius $\delta$ centered at $m(x)$ and $N\{\varepsilon, B_\delta[m(x)], d\}$ be its covering number using balls of size $\varepsilon$. Then
	\[\int_0^1(1+\log N\{\delta\varepsilon, B_{\delta}[m(x)], d\})^{1/2}d\varepsilon=O(1)\quad\mbox{ as }\delta\to0.\]\label{itm:lp3}
	\item
	There exists $\eta_1, \eta_2>0, C_1, C_2>0$ and $\gamma_1, \gamma_2>1$ such that
	\[\inf_{d[m(x), \omega]<\eta_1}\{M(\omega, x)-M[m(x), x]-C_1d[m(x), \omega]^{\gamma_1}\}\geq0,\]
	\[\liminf_{h\to0}\inf_{d[m_{L, h}(x), \omega]<\eta_2}\{M_{L, h}(\omega, x)-M_{L, h}[m_{L, h}(x), x]-C_2d[m_{L, h}(x), \omega]^{\gamma_2}\}\geq0.\]\label{itm:lp4}
\end{enumerate}
Obtaining uniform rates of convergence for local network regression is more involved  and requires stronger conditions. Suppose $\mathcal{T}$ is a closed interval in $\mathbb{R}$. Denote the interior of $\mathcal{T}$ by $\mathcal{T}^o$.
\begin{enumerate}[label=(B\arabic*), leftmargin=1.2cm]
	\setcounter{enumi}{9}
	\item 
	For all $x\in\mathcal{T}$, the minimizers $m(x), m_{L, h}(x)$ and $\hat{m}_{L, n}(x)$ exist and are unique, the last almost surely. Additionally, for any $\varepsilon>0$,
	\[\inf_{x\in\mathcal{T}}\inf_{d[m(x), \omega]>\varepsilon}\{M(\omega, x)-M[m(x), x]\}>0,\]
	\[\liminf_{h\to0}\inf_{x\in\mathcal{T}}\inf_{d[m_{L, h}(x), \omega]>\varepsilon}\{M_{L, h}(\omega, x)-M_{L, h}[m_{L, h}(x), x]\}>0,\]
	and there exists $\zeta=\zeta(\varepsilon)>0$ such that
	\[\p(\inf_{x\in\mathcal{T}}\inf_{d[\hat{m}_{L, n}(x), \omega]>\varepsilon}\{\hat{M}_{L, n}(\omega, x)-\hat{M}_{L, n}[\hat{m}_{L, n}(x), x]\}\geq\zeta)\to1.\]\label{itm:lu2}
	\item
	With $B_\delta[m(x)]\subset\Omega$  and $N\{\varepsilon, B_\delta[m(x)], d\}$ as in Condition \ref{itm:lp3},
	\[\int_0^1\sup_{x\in\mathcal{T}}(1+\log N\{\delta\varepsilon, B_{\delta}[m(x)], d\})^{1/2}d\varepsilon=O(1)\quad\mbox{ as }\delta\to0.\]\label{itm:lu3}
	\item
	There exists $\tau_1, \tau_2>0, D_1, D_2>0$ and $\rho_1, \rho_2>1$ such that
	\[\inf_{x\in\mathcal{T}}\inf_{d[m(x), \omega]<\tau_1}\{M(\omega, x)-M[m(x), x]-D_1d[m(x), \omega]^{\rho_1}\}\geq0,\]
	\[\liminf_{h\to0}\inf_{x\in\mathcal{T}}\inf_{d[m_{L, h}(x), \omega]<\tau_2}\{M_{L, h}(\omega, x)-M_{L, h}[m_{L, h}(x), x]- D_2d[m_{L, h}(x), \omega]^{\rho_2}\}\geq0.\]\label{itm:lu4}
\end{enumerate}
\subsection*{B.2 Proof of Proposition 1}
Each $L\in\mathcal{L}_m$ has the following properties. 
\begin{enumerate}[label=(P\arabic*), leftmargin=1cm]
	\item $L^\T=L$.\label{itm:p0}
	\item The entries in each row sum to $0$, $L1_m=0_m$.\label{itm:p1}
	\item The off-diagonal entries are nonpositive and bounded below, $-W\leq l_{ij}\leq0$.\label{itm:p2}
\end{enumerate}
Properties \ref{itm:p0} and  \ref{itm:p1} can be decomposed into $m(m-1)/2$ and $m$ constraints, respectively. Thus, the dimension of the space of $m\times m$ matrices with Properties \ref{itm:p0} and \ref{itm:p1} is $m^2-m(m-1)/2-m=m(m-1)/2$, and  any matrix satisfying Properties \ref{itm:p0} and \ref{itm:p1} is fully determined by its upper (or lower) triangular submatrix. It is easy to verify that Properties \ref{itm:p0} and  \ref{itm:p1} remain valid under matrix addition and scalar multiplication. Additionally, the matrix consisting of zeros satisfies Properties \ref{itm:p0} and  \ref{itm:p1}. Thus, the space of $m\times m$ matrices with Properties \ref{itm:p0} and \ref{itm:p1} is a subspace of $\mathbb{R}^{m^2}$ of dimension $m(m-1)/2$ and  
$\mathcal{L}_m$ can be bijectively mapped to the hypercube $\{(x^1, \ldots, x^{m(m-1)/2}): -W\leq x^i\leq0\}$, which is bounded, closed, and convex. This proves that $\mathcal{L}_m$ is a bounded, closed, and convex subset in $\mathbb{R}^{m^2}$ of dimension $m(m-1)/2$.
\subsection*{B.3 Proof of Theorem 2 and Theorem 3}
Substituting for the response object $Y$ in Appendix B.1 the $m\times m$ graph Laplacian $L$, which resides in $\mathcal{L}_m$, endowed with the Frobenius metric $d_F$, we show that the metric space $(\mathcal{L}_m, d_F)$ satisfies Conditions \ref{itm:gp0}--\ref{itm:lu4}. Let $\langle\cdot, \cdot\rangle_F$ be the Frobenius inner product. Define 
\begin{align*}
	&B(x)=E(L|X=x);\\&
	B_{G}(x)=E[s_{G}(x)L],\quad\hat{B}_{G}(x)=n^{-1}\sum_{k=1}^ns_{kG}(x)L_k;\\&
	B_{L, h}(x)=E[s_{L}(x, h)L],\quad\hat{B}_{L, n}(x)=n^{-1}\sum_{k=1}^ns_{kL}(x, h)L_k,
\end{align*}
where the expectations and sums for graph Laplacians are element-wise. Since
\begin{align*}
	M(\omega, x)&=E[d_F^2(L, \omega)\mid X=x]\\&=E\{d_F^2[L, B(x)]+d_F^2[B(x), \omega]+2\langle L-B(x), B(x)-\omega\rangle_F\mid X=x\}\\&=M[B(x), x]+d_F^2[B(x), \omega]+2E[\langle L-B(x), B(x)-\omega\rangle_F\mid X=x]
\end{align*}
and
\begin{align*}
	E[\langle L-B(x), B(x)-\omega\rangle_F\mid X=x]&=\langle E(L\mid X=x)-B(x), B(x)-\omega\rangle_F\\&=\langle B(x)-B(x), B(x)-\omega\rangle_F\\&=0,
\end{align*}
one has 
\[M(\omega, x)=M[B(x), x]+d_F^2[B(x), \omega],\]
whence 
\[m(x)=\argmin_{\omega\in\mathcal{L}_m}M(\omega, x)=\argmin_{\omega\in\mathcal{L}_m}d_F^2[B(x), \omega].\]
Additionally, in view of
\begin{align*}
	&E[s_{G}(x)]=1,\quad\frac{1}{n}\sum_{k=1}^ns_{kG}(x)=1,\\&
	E[s_{L}(x, h)]=1,\quad\frac{1}{n}\sum_{k=1}^ns_{kL}(x, h)=1,
\end{align*}
one can similarly show that
\begin{align*}
	&m_{G}(x)=\argmin_{\omega\in\mathcal{L}_m}M_G(\omega, x)=\argmin_{\omega\in\mathcal{L}_m}d_F^2[B_{G}(x), \omega],\\&
	\hat{m}_{G}(x)=\argmin_{\omega\in\mathcal{L}_m}\hat{M}_{G}(\omega, x)=\argmin_{\omega\in\mathcal{L}_m}d_F^2[\hat{B}_{G}(x), \omega],\\&
	m_{L, h}(x)=\argmin_{\omega\in\mathcal{L}_m}M_{L, h}(\omega, x)=\argmin_{\omega\in\mathcal{L}_m}d_F^2[B_{L, h}(x), \omega],\\&
	\hat{m}_{L, n}(x)=\argmin_{\omega\in\mathcal{L}_m}\hat{M}_{L, n}(\omega, x)=\argmin_{\omega\in\mathcal{L}_m}d_F^2[\hat{B}_{L, n}(x), \omega].
\end{align*}
Then by the convexity and closedness of $\mathcal{L}_m$, all the minimizers $m(x)$, $m_{G}(x)$, $\hat{m}_{G}(x)$, $m_{L, h}(x)$, and $\hat{m}_{L, n}(x)$ exist and are unique for any $x\in\mathbb{R}^p$ \citep[chap.~3]{deutsch2012best}. Hence Conditions \ref{itm:gp0}, \ref{itm:gu0} and \ref{itm:lp2}, \ref{itm:lu2} are satisfied. 

To prove that Conditions \ref{itm:gp2}, \ref{itm:gu2} and \ref{itm:lp4}, \ref{itm:lu4} hold, we note that $m(x)$, viewed as the best approximation of $B(x)$ in $\mathcal{L}_m$, is characterized by \citep[chap.~4]{deutsch2012best}
\[\langle B(x)-m(x), \omega-m(x)\rangle_F\leq0,\quad\text{for all}\;\omega\in\mathcal{L}_m.\]
It follows that
\begin{align*}
	M(\omega, x)&=E[d_F^2(L, \omega)\mid X=x]\\&=M[m(x), x]+d_F^2[m(x), \omega]+2E[\langle L-m(x), m(x)-\omega\rangle_F\mid X=x]\\&=M[m(x), x]+d_F^2[m(x), \omega]+2\langle B(x)-m(x), m(x)-\omega\rangle_F\\&\geq M[m(x), x]+d_F^2[m(x), \omega]
\end{align*}
for all $\omega\in\mathcal{L}_m$. Similarly, 
\begin{align*}
	M_G(\omega, x)\geq M_G[m_{G}(x), x]+d_F^2[m_{G}(x), \omega],\\
	M_{L, h}(\omega, x)\geq M_{L, h}[m_{L, h}(x), x]+d_F^2[m_{L, h}(x), \omega],
\end{align*}
for all $\omega\in\mathcal{L}_m$. Consequently, we may select $\eta_i$ and $\tau_i$ arbitrary, $C_i=D_i=1$ and $\gamma_i=\rho_i=2$ for $i=0, 1, 2$ in Conditions \ref{itm:gp2}, \ref{itm:gu2}, and \ref{itm:lp4}, \ref{itm:lu4}. 

Next, we show that Condition \ref{itm:gu1} holds, which then  implies Condition \ref{itm:gp1}. Since $\mathcal{L}_m$ is a subset of $\mathbb{R}^{m^2}$, for any $\omega\in\mathcal{L}_m$ we have
\[N[\delta\varepsilon, B_\delta(\omega), d_F]=N[\varepsilon, B_1(\omega), d_F]\leq (1+2/\varepsilon)^{m^2}.\]
Thus, the integral in Condition \ref{itm:gu1} is bounded by
\begin{align*}
	\int_0^1[1+m^2\log(1+2/\varepsilon)]^{1/2}d\varepsilon&\leq1+m\int_0^1[\log(1+2/\varepsilon)]^{1/2}d\varepsilon\\&\leq1+m\int_0^1[\log(3/\varepsilon)]^{1/2}d\varepsilon\\&=1+3m\int_{\log3}^\infty y^{1/2}e^{-y}dy<\infty,
\end{align*}
using the substitution $y=\log(3/\varepsilon)$. Since this bound does not depend on $\delta$, Condition \ref{itm:gu1} holds and thus Condition \ref{itm:gp1} as well. Likewise we can show that Conditions \ref{itm:lu3} and \ref{itm:lp3} also hold.

Theorem 2 in \citet{mull:19:3} yields rates of convergence for the global network regression estimator. For the local network regression estimator, rates of convergence can be obtained using Corollary 1 in \citet{mull:19:3} and Theorem 1 in \citet{chen2020uniform}. 

\subsection*{B.4 Proof of Proposition 4}
Recall that the matrix power map $F_{\alpha}$ is 
\[F_{\alpha}(S)=S^{\alpha}=U\Lambda^\alpha U^\T: \mathcal{S}_m^+\mapsto\mathcal{S}_m^+,\]
where $U\Lambda U^\T$ is the spectral decomposition of $S$. Specifically, denote the eigenvalues of $S$ by $\lambda_1\geq\lambda_2\geq\cdots\geq\lambda_m\geq0$.  Then 
$F_{\alpha}(S)=U\mathrm{diag}(\lambda_1^\alpha, \lambda_2^\alpha, \ldots, \lambda_m^\alpha)U^\T.$
Note that the power function $f(x)=x^{\alpha}:[0,\infty)\mapsto[0, \infty)$ is $(\alpha, 1)$-H\"{o}lder continuous in $[0, \infty)$ for $0<\alpha<1$, and is $\alpha C^{\alpha-1}$-Lipschitz continuous in $[0, C]$ for $\alpha\geq1$. Results follow directly from Theorem 1.1 in \citet{wihler2009holder} by choosing the scalar function as the power function $f(x)=x^{\alpha}$ with $0<\alpha<1$ in $[0, \infty)$, and with $\alpha\geq1$ in $[0, C]$, respectively.

\subsection*{B.5 Proof of Theorem 5 and Theorem 6}
Substituting for the response object $Y$ in Appendix B.1 a $m\times m$ bounded symmetric positive semi-definite matrix $S$, which resides in $\mathcal{M}_m$, endowed with the Frobenius metric $d_F$, one can show that the metric space $(\mathcal{M}_m, d_F)$ satisfies Conditions \ref{itm:gp0}--\ref{itm:lu4} since $\mathcal{M}_m$, similar to $\mathcal{L}_m$, is a bounded, closed, and convex subset in $\mathbb{R}^{m^2}$. According to Theorem 2 in \citet{mull:19:3}, for the metric space $(\mathcal{M}_m, d_F)$ it holds for $m_G^\alpha(x)$ and $\hat{m}_G^\alpha(x)$ that for a fixed $x\in\mathbb{R}^p$,
\[d_F[m_G^\alpha(x), \hat{m}_G^\alpha(x)]=O_p(n^{-1/2})\]
and for a given $B>0$,
\[\sup_{\|x\|_E\leq B}d_F[m_G^\alpha(x), \hat{m}_G^\alpha(x)]=O_p(n^{-1/[2(1+\varepsilon)]}),\]
for any $\varepsilon>0$.

Next, we  derive rates of convergence when  applying  the inverse matrix power map $F_{1/\alpha}$ and a projection $P_{\mathcal{L}_m}$ onto $\mathcal{L}_m$. Here we  consider two cases:
\begin{enumerate}
	\item Case $0<\alpha\leq1$.\\
	
	\noindent By Proposition 2, 
	\[\|F_{1/\alpha}(S_1)-F_{1/\alpha}(S_2)\|_F\leq\frac{1}{\alpha}(D^\alpha)^{1/\alpha-1}\|S_1-S_2\|_F=\frac{1}{\alpha}D^{1-\alpha}\|S_1-S_2\|_F\]
	for any $S_1, S_2\in\mathcal{M}_m$. Hence for a fixed $x\in\mathbb{R}^p$ 
	\[d_F\{F_{1/\alpha}[m_G^\alpha(x)], F_{1/\alpha}[\hat{m}_G^\alpha(x)]\}=O_p(n^{-1/2})\]
	and for a given $B>0$
	\[\sup_{\|x\|_E\leq B}d_F\{F_{1/\alpha}[m_G^\alpha(x)], F_{1/\alpha}[\hat{m}_G^\alpha(x)]\}=O_p(n^{-1/[2(1+\varepsilon)]}),\]
	for any $\varepsilon>0$.
	
	As shown in the proof for Result 2 in \citet{severn2019manifold}, the projection $P_{\mathcal{L}_m}$ does not increase the distance between two matrices. That is
	\[d_F(P_{\mathcal{L}_m}\{F_{1/\alpha}[m_G^\alpha(x)]\}, P_{\mathcal{L}_m}\{F_{1/\alpha}[\hat{m}_G^\alpha(x)]\})\leq d_F\{F_{1/\alpha}[m_G^\alpha(x)], F_{1/\alpha}[\hat{m}_G^\alpha(x)]\}.\]
	Therefore, for $m_G(x)$ and $\hat{m}_G(x)$ and a fixed $x\in\mathbb{R}^p$,
	\[d_F[m_G(x), \hat{m}_G(x)]=O_p(n^{-1/2})\]
	and for a given $B>0$,
	\[\sup_{\|x\|_E\leq B}d_F[m_G(x), \hat{m}_G(x)]=O_p(n^{-1/[2(1+\varepsilon)]}),\]
	for any $\varepsilon>0$.\\
	
	\item Case $\alpha>1$.\\
	
	By Proposition 2, it holds that
	\[\|F_{1/\alpha}(S_1)-F_{1/\alpha}(S_2)\|_F\leq m^{(\alpha-1)/(2\alpha)}\|S_1-S_2\|_F^{1/\alpha}\]
	for any $S_1, S_2\in\mathcal{M}_m$. Hence we have for a fixed $x\in\mathbb{R}^p$,
	\[d_F\{F_{1/\alpha}[m_G^\alpha(x)], F_{1/\alpha}[\hat{m}_G^\alpha(x)]\}=O_p(n^{-1/(2\alpha)})\]
	and for a given $B>0$,
	\[\sup_{\|x\|_E\leq B}d_F\{F_{1/\alpha}[m_G^\alpha(x)], F_{1/\alpha}[\hat{m}_G^\alpha(x)]\}=O_p(n^{-1/[2(1+\varepsilon)\alpha]}),\]
	for any $\varepsilon>0$.
	
	By the same argument as in the first case, 
	at a fixed $x\in\mathbb{R}^p$,
	\[d_F[m_G(x), \hat{m}_G(x)]=O_p(n^{-1/(2\alpha)}),\]
	and for a given $B>0$,
	\[\sup_{\|x\|_E\leq B}d_F[m_G(x), \hat{m}_G(x)]=O_p(n^{-1/[2(1+\varepsilon)\alpha]}),\]
	for any $\varepsilon>0$.
\end{enumerate}
Similar arguments apply for the local network regression by combining Proposition 2, Corollary 1 in \citet{mull:19:3}, and Theorem 1 in \citet{chen2020uniform}. 

\section*{Appendix C. Additional Simulations}
\label{supp:simu}
\subsection*{C.1 Comparisons Between Different Types of Regression and Metrics}
We replicated simulation scenarios I and III in Section 5 of the paper, corresponding to global and local settings, using both global and local network regression and both  the Frobenius and square root metrics. Specifically, for each simulation scenario, the following four combinations were included: global network regression using the Frobenius metric (GF), global network regression using the square root metric (GS), local network regression using the Frobenius metric (LF) and  local network regression using the square root metric (LS). The corresponding mean integrated squared errors are summarized in Table~\ref{tab:simucomresults}. Under simulation scenario I, where each entry of the graph Laplacian $L$ is conditionally linear in the predictor $X$, the global network regression using the Frobenius metric is seen to perform best, validating the superiority of global approaches if the linear assumption holds. In contrast, if there is an underlying nonlinear relationship as in simulation scenario III, local network regression leads to much smaller errors, which is as expected. We also note  that the square root metric achieves better performance  under simulation scenario I with local linear regression. 
\begin{table}[t]
	\centering
	\begin{tabular}{l|cccc|cccc}
		\hline
		Sample size & \multicolumn{4}{c|}{I} & \multicolumn{4}{c}{III}\\\hline
		& GF & GS & LF & LS & GF & GS & LF & LS\\
		50 & 0.467 & 4.025 & 7.116 & 1.348 & 81.295 & 87.481 & 1.557 & 2.227\\
		100 & 0.233 & 3.741 & 3.621 & 0.981 & 83.612 & 88.966 & 0.860 & 1.458\\
		200 & 0.117 & 3.664 & 2.261 & 0.732 & 84.327 & 89.215 & 0.486 & 0.998\\
		500 & 0.047 & 3.588 & 1.090 & 0.542 & 84.938 & 89.539 & 0.230 & 0.644\\
		1000 & 0.024 & 3.569 & 0.613 & 0.468 & 85.103 & 89.644 & 0.132 & 0.497\\\hline
	\end{tabular}
	\caption{Mean integrated squared errors for five sample sizes under simulation scenarios I and III in Section 5 of the paper using the proposed methods (GF, global network regression using the Frobenius metric; GS, global network regression using the square root metric; LF, local network regression using the Frobenius metric; LS, local network regression using the square root metric).}
	\label{tab:simucomresults}
\end{table}

\subsection*{C.2 Networks Generated from Erd\H{o}s-R\'{e}nyi Model}
Here we assess the performance of global and local network regression estimates on networks generated from Erd\H{o}s-R\'{e}nyi random graph model \citep{erdHos1959random}. Similar to Section 5,  four different simulation scenarios I - IV are considered, corresponding to global and local network regression using both the Frobenius metric $d_F$ and the square root metric $d_{F, 1/2}$. Specifically, consider the Erd\H{o}s-R\'{e}nyi random graph model $G(m, N)$, where networks are  sampled uniformly at random from the collection of all networks which have $m$ nodes and $N$ edges. The probability of the presence of each edge is thus $N/M$ where $M = m(m - 1)/2$. The weights of existing edges are assumed to follow a beta distribution with shape parameters $\alpha = X$ and $\beta = 1-X$ for global network regression or $\alpha = \sin(\pi X)$ and $\beta = 1-\sin(\pi X)$ for local network regression. The associated graph Laplacian is then taken as the random response $L$ for the proposed regression models. In particular for global network regression under the square root metric, the random response is taken as $F_2(L)$ to ensure that the linearity assumption is satisfied.

\begin{figure}[tb]
	\centering
	\includegraphics[width=0.7\linewidth]{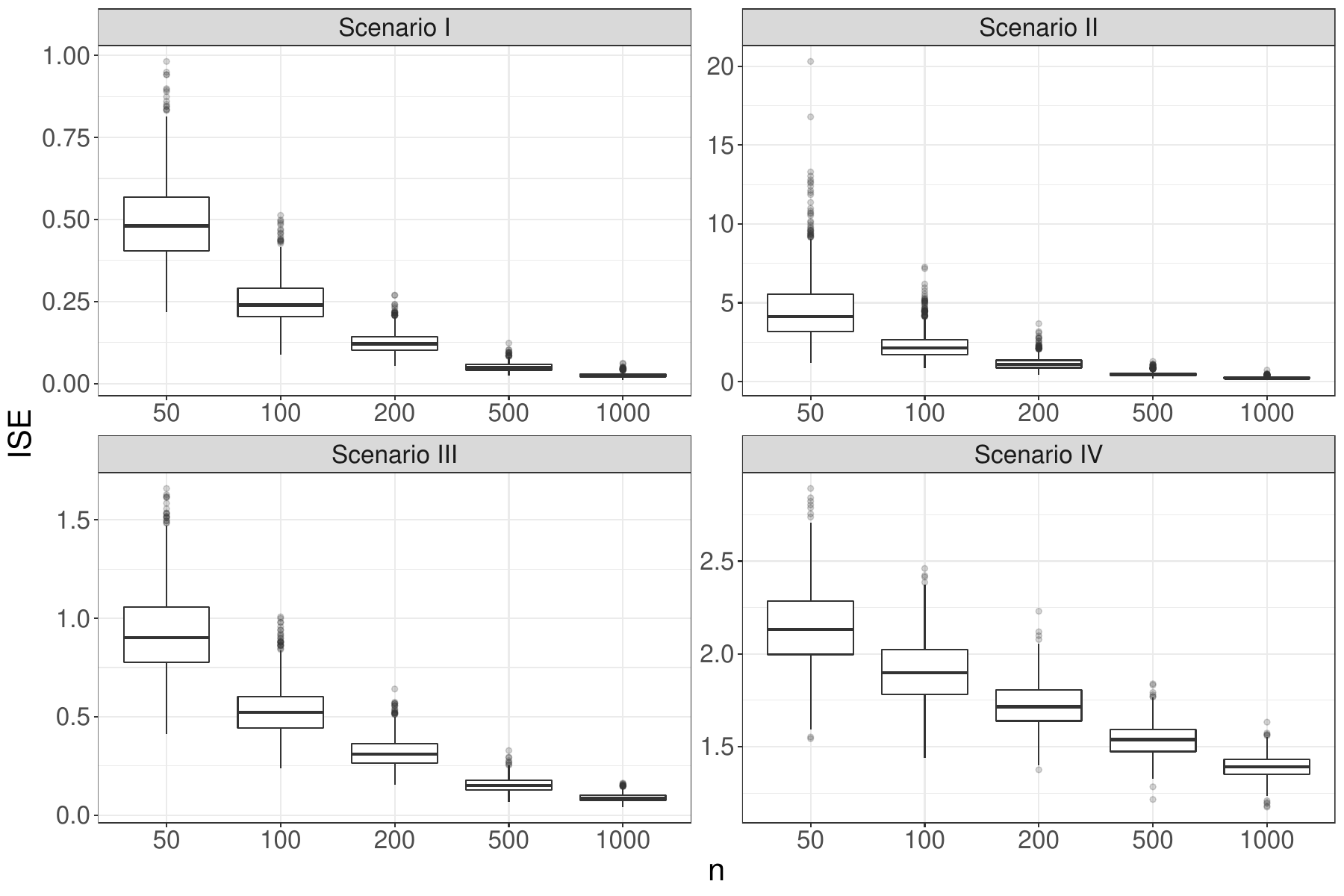}
	\caption{Boxplots of integrated square errors (ISE) for networks generated from Erd\H{o}s-R\'{e}nyi random graph model.}
	\label{fig:simuer}
\end{figure}

We investigated sample sizes $n=50, 100, 200, 500, 1000$, with $Q=1000$ Monte Carlo runs for each simulation scenario. In each iteration, random samples of pairs $(X_k, L_k)$, $k=1, \ldots, n$ were generated by sampling $X_k\sim\mathrm{U}(0, 1)$, setting $m=10$, $N=9$, and following the above procedure. The quality of the estimation for each simulation run was quantified by the integrated squared error as defined in Section 5. The bandwidths for the local network regression were chosen by leave-one-out cross-validation. The integrated squared errors (ISE) for $Q=1000$ simulation runs and five sample sizes under four simulation scenarios are summarized in the boxplots in Figure~\ref{fig:simuer}. With increasing sample size, the integrated squared errors are seen to decrease, demonstrating the validity and utility of the proposed network regression models for various network generative models.

\section*{Appendix D. Additional Materials}
\label{supp:add}

\begin{figure}[tb]
	\centering
	\begin{subfigure}{.45\textwidth}
		\centering
		\includegraphics[width=\linewidth]{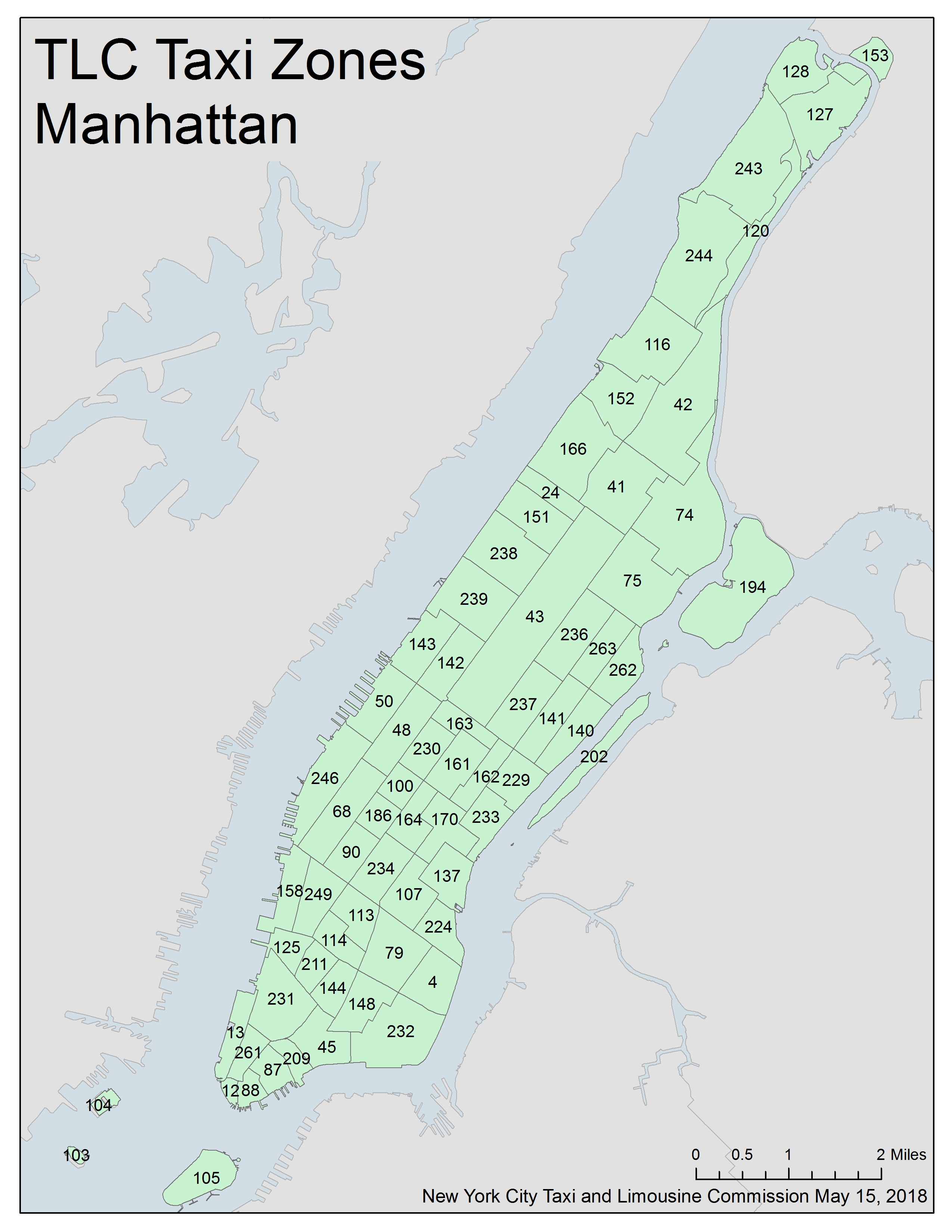}
		\caption{}
	\end{subfigure}%
	\begin{subfigure}{.45\textwidth}
		\centering
		\includegraphics[width=\linewidth]{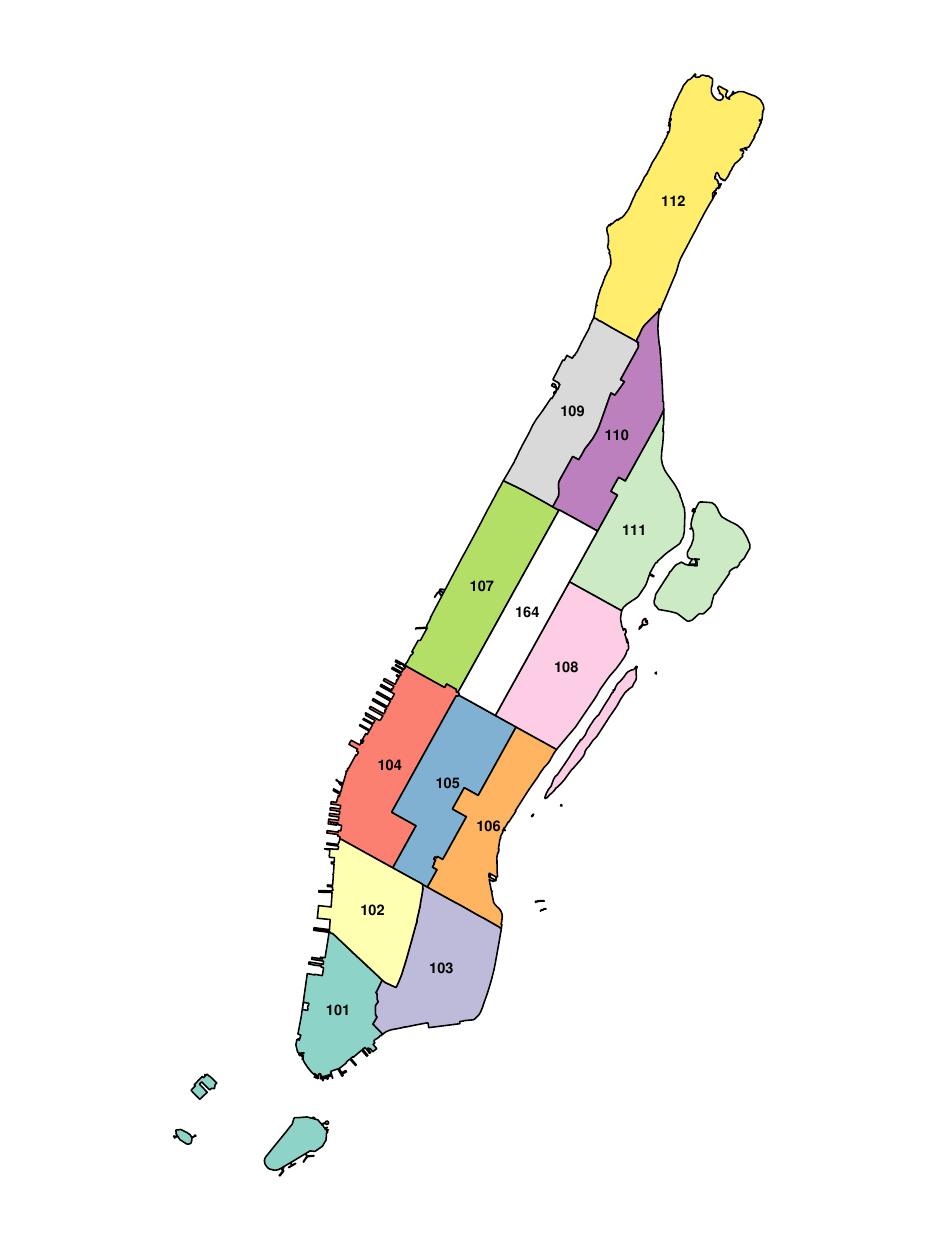}
		\caption{}
	\end{subfigure}
	\caption{(a) TLC taxi zones in Manhattan, New York. (b) $13$ regions in Manhattan, New York.}
	\label{fig:taxizone}
\end{figure}

The yellow and green taxi trip records on pick-up and drop-off dates/times, pick-up and drop-off locations, trip distances, itemized fares, rate types, payment types, and driver-reported passenger counts, collected by the New York City Taxi and Limousine Commission (NYC TLC), are publicly available at \url{https://www1.nyc.gov/site/tlc/about/tlc-trip-record-data.page}. The taxi zone map for Manhattan (Figure~\ref{fig:taxizone}(a)) available at \url{https://www1.nyc.gov/assets/tlc/images/content/pages/about/taxi_zone_map_manhattan.jpg} represents the boundaries zones for taxi pick-ups and drop-offs as delimited by the NYC TLC. We excluded the islands ($103$ Liberty Island, $104$ Ellis Island, $105$ Governor's Island) from our study. The remaining $66$ zones in Manhattan can be grouped into $13$ regions as delimited in Figure~\ref{fig:taxizone}(b) (source: \url{https://communityprofiles.planning.nyc.gov}). For the composition of these $13$ regions see Table~\ref{table:zones}.

\begin{table}[tb]
	\centering
	\begin{tabularx}{\textwidth}{|l||X|}
		\hline
		Region & Zones \\ \hline\hline
		$101$  & 12 Battery Park, 13 Battery Park City, 87 Financial District North, 88 Financial District South, 209 Seaport, 231 TriBeCa/Civic Center, 261 World Trade Center \\ \hline
		$102$  & 113 Greenwich Village North, 114 Greenwich Village South, 125 Hudson Sq, 144 Little Italy/NoLiTa, 158 Meatpacking/West Village West, 211 SoHo, 249 West Village \\ \hline
		$103$  & 4 Alphabet City, 45 Chinatown, 79 East Village, 148 Lower East Side, 232 Two Bridges/Seward Park \\ \hline
		$104$  & 48 Clinton East, 50 Clinton West, 68 East Chelsea, 90 Flatiron, 246 West Chelsea/Hudson Yards\\ \hline
		$105$  & 100 Garment District, 161 Midtown Center, 163 Midtown North, 164 Midtown South, 186 Penn Station/Madison Sq West, 230 Times Sq/Theatre District, 234 Union Sq \\ \hline
		$106$  & 107 Gramercy, 137 Kips Bay, 162 Midtown East, 170 Murray Hill, 224 Stuy Town/Peter Cooper Village, 229 Sutton Place/Turtle Bay North, 233 UN/Turtle Bay South \\ \hline
		$107$  & 24 Bloomingdale, 142 Lincoln Square East, 143 Lincoln Square West, 151 Manhattan Valley, 238 Upper West Side North, 239 Upper West Side South \\ \hline
		$108$  & 140 Lenox Hill East, 141 Lenox Hill West, 202 Roosevelt Island, 236 Upper East Side North, 237 Upper East Side South, 262 Yorkville East, 263 Yorkville West \\ \hline
		$109$  & 116 Hamilton Heights, 152 Manhattanville, 166 Morningside Heights \\ \hline
		$110$ & 41 Central Harlem, 42 Central Harlem North \\ \hline
		$111$  & 74 East Harlem North, 75 East Harlem South, 194 Randalls Island \\ \hline
		$112$  & 120 Highbridge Park, 127 Inwood, 128 Inwood Hill Park, 153 Marble Hill, 243 Washington Heights North, 244 Washington Heights South \\ \hline
		$164$ & 43 Central Park \\ \hline
	\end{tabularx}
	\caption{Details about $13$ regions in Manhattan, New York.}
	\label{table:zones}
\end{table}

The $45$ regions in each hemisphere as delimited by the automated anatomical labeling (AAL) atlas \citep{tzourio2002automated} are listed in Table~\ref{tab:aal}.
\newpage
{\footnotesize
	\begin{longtable}{|lll|}
		\hline
		ROI            & Lobe                                      & Label \\ \hline
		Central region & ~                                         & ~    \\ 
		\hspace{0.5cm}1              & Precentral gyrus                          & PRE  \\ 
		\hspace{0.5cm}2              & Postcentral gyrus                         & POST \\ 
		\hspace{0.5cm}3              & Rolandric operculum                       & RO   \\ 
		Frontal lobe   & ~                                         & ~    \\ 
		\hspace{0.5cm}Lateral surface   & ~                                         & ~    \\ 
		\hspace{0.5cm}4              & Superior frontal gyrus, dorsolateral      & F1   \\ 
		\hspace{0.5cm}5              & Middle frontal gyrus                      & F2   \\ 
		\hspace{0.5cm}6              & Inferior frontal gyrus, opercular part    & F3OP \\ 
		\hspace{0.5cm}7              & Inferior frontal gyrus, triangular part   & F3T  \\ 
		\hspace{0.5cm}Medial surface   & ~                                         & ~    \\ 
		\hspace{0.5cm}8              & Superior frontal gyrus, medial            & F1M  \\ 
		\hspace{0.5cm}9              & Supplementary motor area                  & SMA  \\ 
		\hspace{0.5cm}10             & Paracentral lobule                        & PCL  \\ 
		\hspace{0.5cm}Orbital surface   & ~                                         & ~    \\ 
		\hspace{0.5cm}11             & Superior frontal gyrus, orbital part      & F1O  \\ 
		\hspace{0.5cm}12             & Superior frontal gyrus, medial orbital    & F1MO \\ 
		\hspace{0.5cm}13             & Middle frontal gyrus, orbital part        & F2O  \\ 
		\hspace{0.5cm}14             & Inferior frontal gyrus, orbital part      & F3O  \\ 
		\hspace{0.5cm}15             & Gyrus rectus                              & GR   \\ 
		\hspace{0.5cm}16             & Olfactory cortex                          & OC   \\ 
		Temporal lobe  & ~                                         & ~    \\ 
		\hspace{0.5cm}Lateral surface   & ~                                         & ~    \\ 
		\hspace{0.5cm}17             & Superior temporal gyrus                   & T1   \\ 
		\hspace{0.5cm}18             & Heschl gyrus                              & HES  \\ 
		\hspace{0.5cm}19             & Middle temporal gyrus                     & T2   \\ 
		\hspace{0.5cm}20             & Inferior temporal gyrus                   & T3   \\ 
		Parietal lobe  & ~                                         & ~    \\ 
		\hspace{0.5cm}Lateral surface   & ~                                         & ~    \\ 
		\hspace{0.5cm}21             & Superior parietal gyrus                   & P1   \\ 
		\hspace{0.5cm}22             & Inferior parietal                         & P2   \\ 
		\hspace{0.5cm}23             & Angular gyrus                             & AG   \\ 
		\hspace{0.5cm}24             & Supramarginal gyrus                       & SMG  \\ 
		\hspace{0.5cm}Medial surface   & ~                                         & ~    \\ 
		\hspace{0.5cm}25             & Precuneus                                 & PQ   \\ 
		Occipital lobe & ~                                         & ~    \\ 
		\hspace{0.5cm}Lateral surface   & ~                                         & ~    \\ 
		\hspace{0.5cm}26             & Superior occipital gyrus                  & O1   \\ 
		\hspace{0.5cm}27             & Middle occipital gyrus                    & O2   \\ 
		\hspace{0.5cm}28             & Inferior occipital gyrus                  & O3   \\ 
		\hspace{0.5cm}Medial surface   & ~                                         & ~    \\ 
		\hspace{0.5cm}29             & Cuneus                                    & Q    \\ 
		\hspace{0.5cm}30             & Calcarine Fissure                         & V1   \\ 
		\hspace{0.5cm}31             & Lingual gyrus                             & LING \\ 
		\hspace{0.5cm}32             & Fusiform gyrus                            & FUSI \\ 
		Limbic lobe    & ~                                         & ~    \\ 
		\hspace{0.5cm}33             & Temporal pole: superior temporal gyrus    & T1P  \\ 
		\hspace{0.5cm}34             & Temporal pole: middle temporal gyrus      & T2P  \\ 
		\hspace{0.5cm}35             & Anterior cingulate and paracingulate gyri & ACIN \\ 
		\hspace{0.5cm}36             & Median cingulate and paracingulate gyri   & MCIN \\ 
		\hspace{0.5cm}37             & Posterior cingulate gyrus                 & PCIN \\ 
		\hspace{0.5cm}38             & Hippocampus                               & HIP  \\ 
		\hspace{0.5cm}39             & Parahippocampal gyrus                     & PHIP \\ 
		\hspace{0.5cm}40             & Insula                                    & INS  \\ 
		Subcortical    & ~                                         & ~    \\ 
		\hspace{0.5cm}41             & Amygdala                                  & AMYG \\ 
		\hspace{0.5cm}42             & Caudate nuclei                            & CAU \\
		\hspace{0.5cm}43             & Lenticular nucleus, putamen                     & PUT \\
		\hspace{0.5cm}44             & Lenticular nucleus, pallidum                      & PAL \\
		\hspace{0.5cm}45             & Thalamus                            & THAL \\\hline
		\caption{Anatomical regions of interest (ROIs) in each hemisphere for the AAL atlas.}
		\label{tab:aal}
	\end{longtable}
}

\vskip 0.2in
\bibliography{sample}

\end{document}